\documentclass[twocolumn,english,aps,prl,showpacs]{revtex4}
\usepackage[T1]{fontenc}
\usepackage{amsmath,graphicx,amssymb,epsfig,babel,dsfont,color}

\renewcommand{\Re}{{\rm Re}}
\renewcommand{\Im}{{\rm Im}}
\newcommand{\Tr}{{\rm Tr}}
\newcommand{\rd}{{\rm d}}
\newcommand{\kb}{k_{\rm B}}
\newcommand{\rL}{{\rm L}}
\newcommand{\rR}{{\rm R}}

\newcommand{\rD}{{\rm D}}
\newcommand{\rS}{{\rm S}}
\newcommand{\rG}{{\rm G}}
\newcommand{\rE}{{\rm E}}

\newcommand{\rs}{{\rm s}}
\newcommand{\rp}{{\rm p}}
\newcommand{\rc}{{\rm c}}
\newcommand{\kB}{k_{\rm B}}

\begin{document}

\title{Thermotronics: toward nanocircuits to manage radiative heat flux }

\author{Philippe Ben-Abdallah}
\email{pba@institutoptique.fr}
\affiliation{Laboratoire Charles Fabry,UMR 8501, Institut d'Optique, CNRS, Universit\'{e} Paris-Sud 11,
2, Avenue Augustin Fresnel, 91127 Palaiseau Cedex, France}
\affiliation{Universit\'{e} de Sherbrooke, Department of Mechanical Engineering, Sherbrooke, PQ J1K 2R1, Canada.}

\author{Svend-Age Biehs}
\email{s.age.biehs@uni-oldenburg.de}
\affiliation{Institut f\"{u}r Physik, Carl von Ossietzky Universit\"{a}t, D-26111 Oldenburg, Germany.}

\date{\today}


\date{\today}

\begin{abstract}
The control of electric currents in solids  is  at the origin of the modern electronics revolution which has driven our daily life since the second half of 20th century. Surprisingly, to date, there is no thermal analog for a  control of heat flux. Here, we summarize the very last developments carried out in this direction
to control heat exchanges by radiation both in near and far-field in complex architecture networks. 

\end{abstract}


\maketitle

%
%

\section{Introduction}

The control of electric currents in solids is at the origin of modern electronics which has revolutionized our daily life. The diode and the transistor introduced by Braun~\cite{Braun} and Bardeen~\cite{Bardeen} are undoubtedly the corner stones of modern information technologies. Such devices allow for rectifying, switching, modulating and even amplifying the electric current. Beside these developements, in the mid 1890s, Tesla's work~\cite{Tesla} on its remotely-controlled device also called  "telautomaton"  conceived electronically tunable components with respect to the value of external excitations (electromotrice forces) giving rise to the very first logic gates.

Astonishingly, similar devices which would make possible the control of heat flow do not exist in our current life. An important step forward in this direction has been carried out  in 2006 by Baowen Li and co-workers {\itshape et al.}~\cite{Casati1} by introducing phononic counterpart of a field-effect transistor. In their device, composed as its electronic analog, of three interconnected solid elements, the temperature bias  plays the role of the voltage bias and the heat currents carried by phonons play the role of the electric currents. Later, several prototypes of phononic thermal logic gates~\cite{BaowenLi2} as well as thermal memories~\cite{BaowenLi3,BaowenLiEtAl2012} have been developed in order to process information by phononic heat currents rather than by electric currents. 
Beside these results, different phononic thermal rectifiers have been proposed~\cite{BaowenLi2004,Chang} to introduce an asymmetry in the heat transport with respect to the sign of the temperature gradient~\cite{Starr1936,RobertsWalker2011} opening so the way to the development of thermal diodes.

However, this transport of heat with phonons in solid networks suffers from some weaknesses of fundamental nature which intrinsically limit its performances. One of these limitations is linked to the speed of acoustic phonons itself which is limited by the speed of sound in solids. Another intrinsic limitation of phononic devices is the  presence of local Kapitza resistances which come from the mismatch of vibrational modes supported by the different solid elements in the network. This resistance  can drastically reduce the heat flux transported across the system. 

In this paper we discuss a photonic alternative to the phononic technology by making a review of the latest developments in this direction showing the possibility to realize thermal analogs of electronic fundamental building blocks such as thermal transistor, thermal memory and thermal logic gates for controlling the flow of heat by radiation, storing thermal energy and even making logical operations using thermal photons instead of electrons. 
Finally, we suggest new research directions for advanced thermal management by tuning radiative heat exchanges  in many-body systems using magnetic fields.

%
%

\section{Thermal transistor}

As outlined in the introduction, a purely photonic technology has been proposed as an alternative to the phononic one. An important step forward in this direction has been carried out by Fan et al.~\cite{OteyEtAl2010} by introducing the first radiative thermal rectifier  by exploiting the thermal dependence of optical properties of materials in interaction. During the next five years the rectification performances have been improved~\cite{BasuFrancoeur2011, NefzaouiEtAl2013,Zhang2,Huang,Dames,Zhu2} until the development of a phase-change radiative thermal diode  able to rectify 66\% of heat flux in far-field~\cite{PBA_APL,Ito} and even more than 99\% in the near-field regime~\cite{van Zwol1}, respectively.  Contrary to the relatively weak dependence with respect to the temperature of optical properties of materials used in the thermal rectifier developed  so far, the phase-change thermal diode exploites a sudden change of these properties around a critical temperature, the transition temperature.

\begin{figure}
\includegraphics[scale=0.35,angle=0]{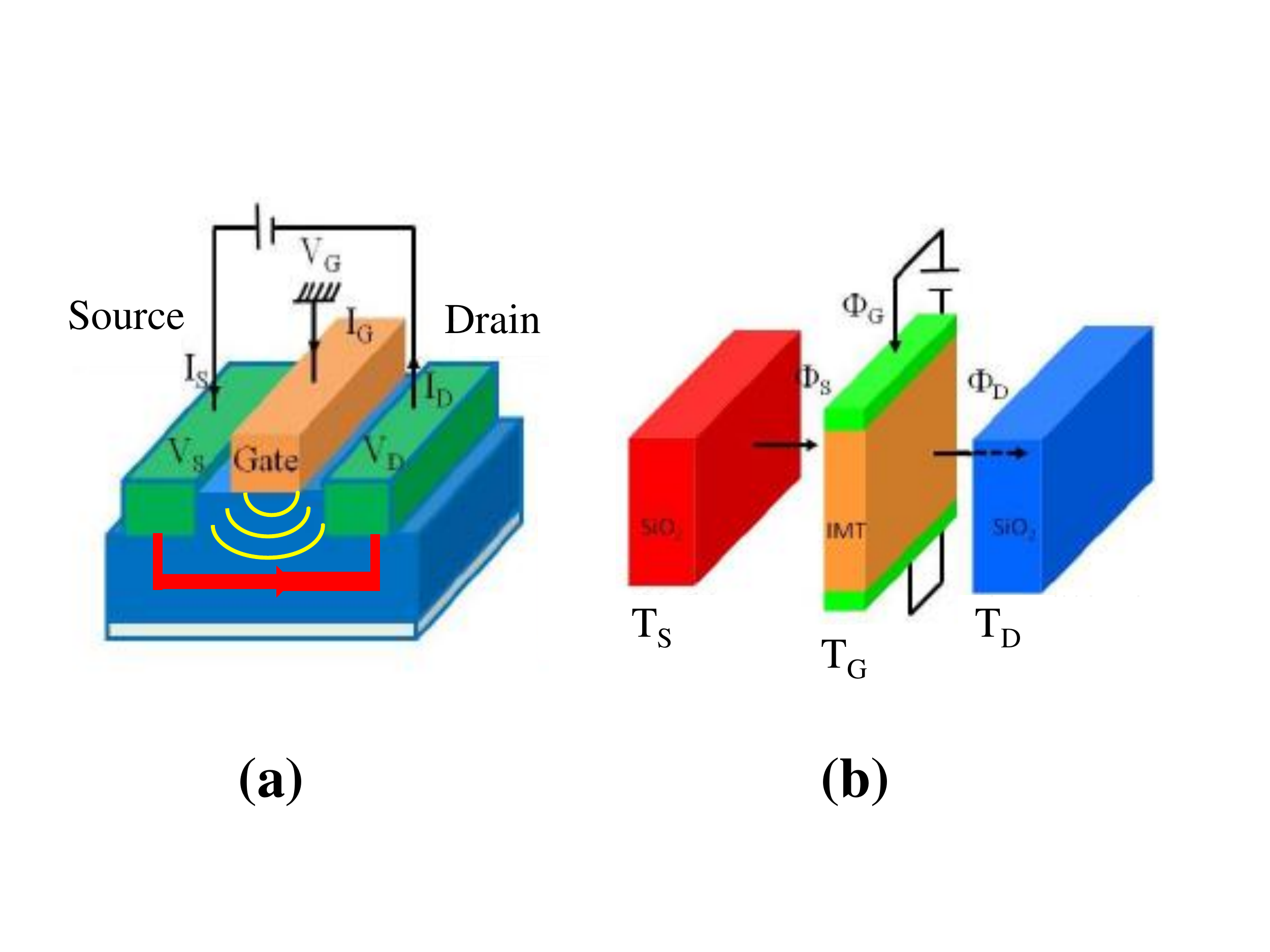} 
\caption{(a) Electronic transistor made of three terminals, the source (S), the gate (G), 
         and the drain (D). The gate is used to actively control (by applying a voltage bias $V$ on it) the apparent 
         electric conductivity of the channel between the source and the drain. (b)
         Radiative thermal transistor. A membrane of a MIT material (VO$_2$) acts as the gate 
         between two silica (SiO$_2$) thermal reservoirs (source and drain). The temperature ($T_\rG$)
         of the gate is chosen between the temperatures of the source and drain 
         ($T_\rS$ and $T_\rD$). $\Phi_\rD$ and $\Phi_\rS$ are the radiative heat fluxes received by the drain and emitted 
         by the source, respectively.}.
\label{transistor}
\end{figure}

Following this idea, a general concept of radiative transistor has been introduced in 2014~\cite{PBA_PRL2014} and applied in 2015~\cite{JoulainAPL2015} in the particular situation where only propagating photons participate to the transfer. Before introducing it, let us briefly describe
the classical electronic transistor. This last is
sketched in Fig.~\ref{transistor}(a). It is composed by three 
solid elements, the drain, the source, and the gate. It is basically used to control the flux of electrons 
exchanged in the channel between the drain and the source by changing the voltage bias applied 
on the gate. While, the physical diameter of this channel is fixed its effective electrical diameter can be 
tuned by application of a voltage on the gate. A tiny change in this voltage can cause a large variation 
in the current from the source to the drain. 

\begin{figure}
\includegraphics[scale=0.3,angle=0]{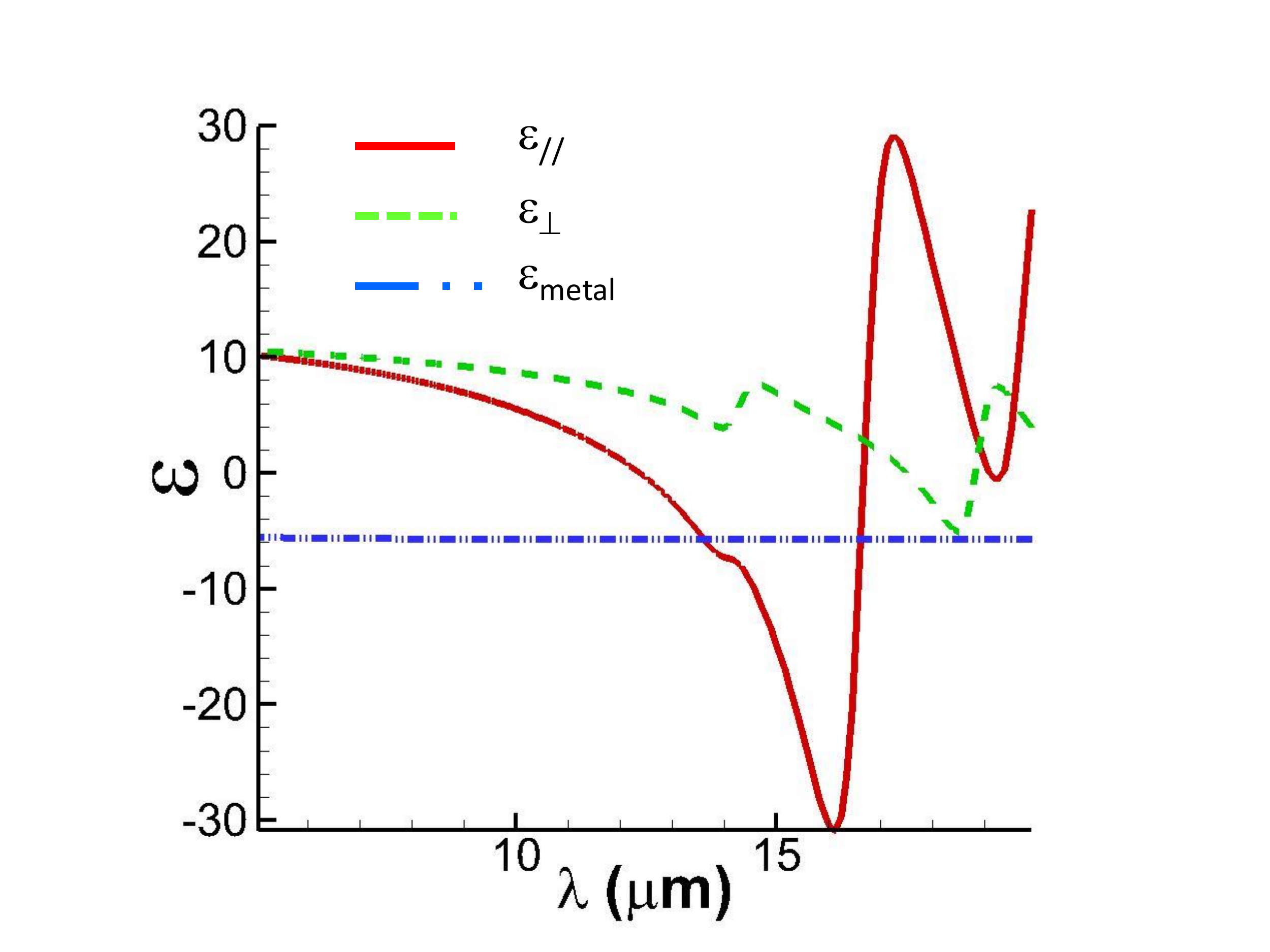} 
\caption{Permittivity (real part) of VO$_2$ in its metallic phase ($\epsilon_{\rm metal}$ when $T>T_c$) and in its cristalline 
          phase ($\epsilon_\parallel$ and $\epsilon_\perp$ when $T<T_c$), $T_c$ being the critical temperature of VO$_2$. \label{Fig:Perm}}.
\label{permittivity}
\end{figure}

The radiative analog of an electronic transistor~\cite{PBA_PRL2014} is depicted in Fig.~\ref{transistor}(b).
It  basically consists in a  source and a drain, labeled by the indices S and D, which are maintained at 
temperatures $T_\rS$ and $T_\rD$ (which play an analog role as the voltage) using thermostats where $T_\rS > T_\rD$ so that  a net heat flux is 
transferred from the source towards the drain. A thin layer of a metal-isulator transition material (MIT)  labeled by G of 
width $\delta$ is placed between the source and the drain at a distance $d$ from both media and operates as a gate. 
This configuration coincides with two heat-radiation diodes~\cite{PBA_APL} which are connected in series, so that
the heat radiation transistor corresponds to a bipolar transistor.
In a MIT material a small change in the temperature around its critical temperature $T_c$ causes a sudden qualitative and quantitative change  in its optical properties. Vanadium dioxide (VO$_2$) is one of 
such materials (see Fig.~\ref{permittivity}) which undergoes a first-order 
transition (Mott transition~\cite{Mott}) from a high-temperature metallic phase to a low-temperature insulating 
phase~\cite{Baker} close to room-temperature ($T_\rc=340\,{\rm K}$). Different works have shown~\cite{van Zwol1,van Zwol2,vanZwol3} that the heat-flux exchanged at 
close separation distances (i.e.\ in the near-field regime) between an MIT material and another medium, can be modulated by several orders of magnitude across 
the phase transition of MIT materials.

Without external excitation, the system will reach its steady state for which the net flux $\Phi_\rG$ 
received by the intermediate medium, the gate, is zero by heating or cooling the gate until it reaches 
its steady state or equilibrium temperature $T_\rG^{\rm eq}$. In this case the gate temperature 
$T_\rG$ is set by the temperature of the surrounding media, i.e.\ the drain and the source. When a certain amount 
of heat is added to or removed from the gate for example by applying a voltage difference through a 
couple of electrodes as illustrated in Fig.~\ref{transistor}(b) or by extracting heat using Peltier 
elements, its temperature can be either increased or reduced around its equilibrium temperature $T_G^{eq}$. 
This external action on the gate allows to tailor the heat flux $\Phi_\rS$ between the source and the gate 
and the heat flux $\Phi_\rD$ between the gate and the drain. 

To show that this system operates as a transistor let us examinate how the radiative heat flux evolves in this system with respect to the gate temperature. In a three-body system the radiative flux received by the drain 
takes the form~\cite{Messina,Messina2}
\begin{equation}
  \Phi_{\rD}= \int_0^\infty\!\frac{d\omega}{2\pi}\,\phi_\rD(\omega,d), 
\label{Eq:Flux_D}
\end{equation}
where the spectral heat flux is given by
\begin{equation}
\begin{split}
  \phi_{\rD} &= \sum_{j = \{\rm s,p\}}\int\! \frac{{\rm d}^2 \boldsymbol{\kappa}}{(2 \pi)^2} \, 
                \bigl[\Theta_{\rS\rG}(\omega)\mathcal{T}^{\rS/\rG}_j(\omega,\boldsymbol{\kappa}; d)\\
                &\quad+ \Theta_{\rm GD}(\omega)\mathcal{T}^{\rG/\rD}_j(\omega,\boldsymbol{\kappa}; d)\bigr].
\end{split}
\label{Eq:Flux_D_1}
\end{equation}
Here, $\mathcal{T}^{\rS/\rG}_j$ and $\mathcal{T}^{\rG/\rD}_j$ denote the transmission coefficients of each mode $(\omega,\boldsymbol{\kappa})$ 
between the source and the gate and between the gate and the drain for both polarization states $j=\rs,\rp$. In the above relation 
$\Theta_{ij}$ denotes the difference of functions $\Theta(\omega,T_i)$ and $\Theta(\omega,T_j)$, $\Theta(\omega,T)=\frac{\hbar\omega}{ (\exp\bigl({\frac{\hbar \omega}{\kB T}}\bigr)-1)}$ being the mean energy of a Planck oscillator at temperature $T$. 
According to the N-body near-field heat transfer theory~\cite{Messina,Messina2}, the transmission coefficients $\mathcal{T}^{\rS/\rG}_j$ and 
$\mathcal{T}^{\rG/\rD}_j$ of the energy carried by each mode written in terms of optical reflection coefficients $\rho_{E,j}$ ($\rE = \rS, \rD, \rG$) 
and transmission coefficients $\tau_{E,j}$ of each basic element of the system and in terms of reflection coefficients 
$\rho_{EF,j}$ ($E = \rS, \rD, \rG$ and $F = \rS, \rD, \rG$) of couples of elementary elements is given by ($\kappa > \omega/c$)
\begin{equation}
\begin{split}
  &\mathcal{T}^{\rS/\rG}_{j}(\omega,\boldsymbol{\kappa},d)\\
  &=\frac{4\mid\tau_{\rG,j}\mid^2 \Im(\rho_{\rS,j})\Im(\rho_{\rD,j})e^{-4\gamma d}}{\mid 1-\rho_{\rS\rG,j}\rho_{\rD,j}e^{-2\gamma d}\mid^2\mid1-\rho_{\rS,j}\rho_{\rG,j}e^{-2\gamma d}\mid^2},\\
  &\mathcal{T}^{\rG/\rD}_{j}(\omega,\boldsymbol{\kappa},d)=\frac{4 \Im(\rho_{\rS\rG,j})\Im(\rho_{\rD,j})e^{-2\gamma d}}{\mid 1-\rho_{\rS\rG,j}\rho_{\rD,j}e^{-2\gamma d}\mid^2}
\end{split}
\label{Trans1}
\end{equation}
introducing the imaginary part of the wavevector normal to the surfaces in the multilayer structure $\gamma = \Im(k_{z0})$. 
Similarly the heat flux from the source towards the gate reads
\begin{equation}
\begin{split}
  \phi_{\rS} &= \sum_{j = \{\rm s,p\}}\int\! \frac{{\rm d}^2 \boldsymbol{\kappa}}{(2 \pi)^2} \, \bigl[\Theta_{\rD\rG}(\omega)\mathcal{T}^{\rD/\rG}_j(\omega,\boldsymbol{\kappa}; d)\\
             &\quad + \Theta_{\rG\rS}(\omega)\mathcal{T}^{\rG/\rS}_j(\omega,\boldsymbol{\kappa}; d)\bigr]
\end{split}
\label{Eq:Flux_S_1}
\end{equation}
where the transmission coefficients are analog to those defined in Eq.~(\ref{Trans1}) and can be obtained making the substitution $\rS \leftrightarrow \rD$. Note that here we have neglected the contribution of the propagating waves which is vanishingly small compared to the contribution of the evanescent waves for distances $d$ much smaller than the thermal wavelength. At steady state, the net heat flux received/emitted by the gate which is just given by the heat flux from the source to the gate minus the heat flux from the gate to the drain vanishes so that
\begin{equation}
 \Phi_\rG = \Phi_\rS - \Phi_\rD = 0.
\label{Eq:equilibrium}
\end{equation}
This relation allows us to identify the gate equilibrium temperature $T^{\rm eq}_\rG$ (which is not necessarily unique because of 
the presence of bistability mechanisms~\cite{PBA_PRL2014}) for given temperatures $T_\rS$ and $T_\rD$. Note that out of steady state, 
the heat flux received/emitted by the gate is $\Phi_\rG = \Phi_\rS - \Phi_\rD \neq 0$. If $\Phi_\rG < 0$ ($\Phi_\rG > 0$) an external 
flux is added to (removed from) the gate by heating (cooling). 

\begin{figure}
\includegraphics[scale=0.35,angle=0]{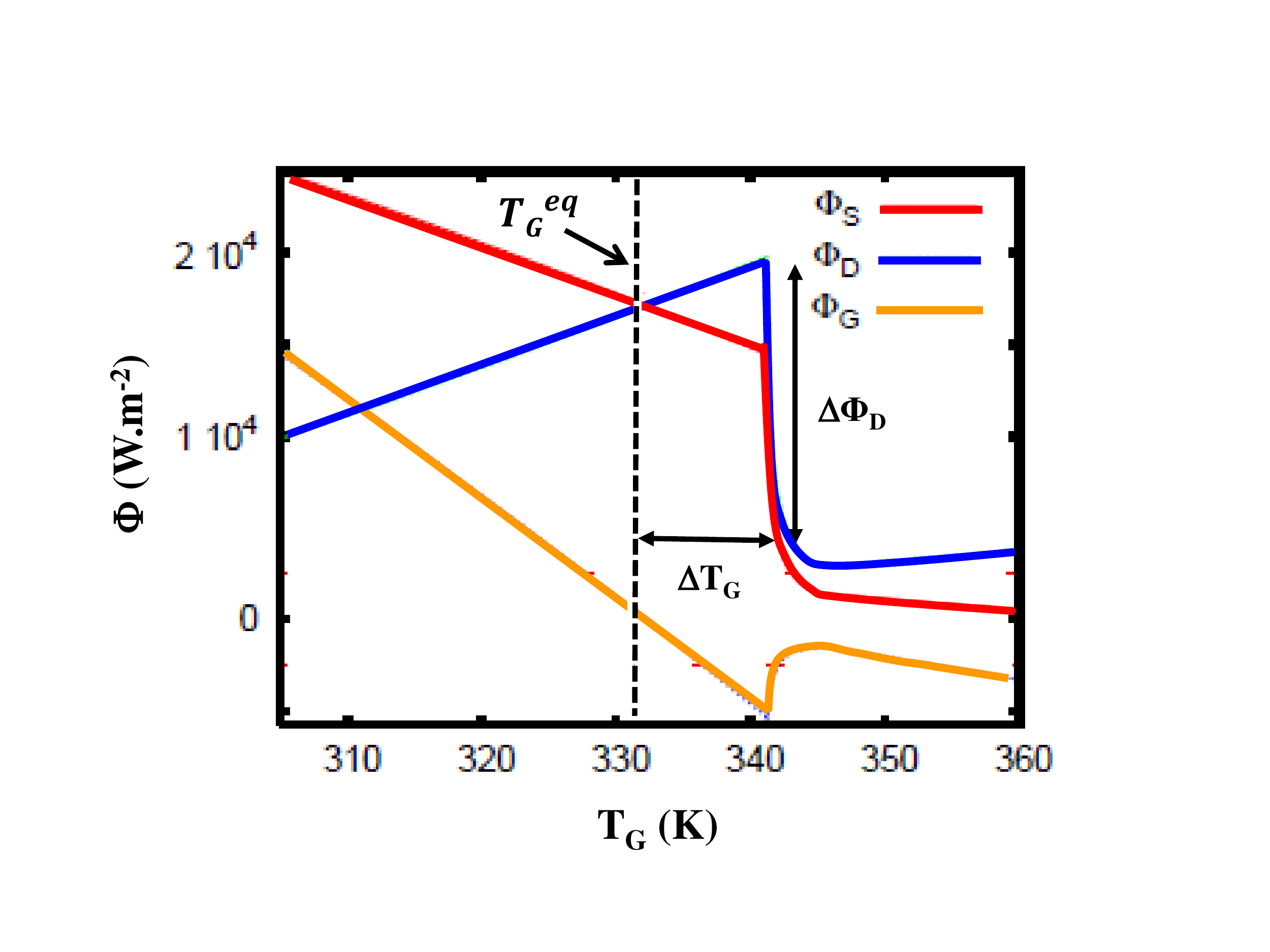} 
\caption{Heat fluxes inside the near-field thermal transistor with a  
         VO$_2$ gate of thickness $\delta= 50\,{\rm nm}$ located at a distance 
         $d = 100\,{\rm nm}$ between two massive silica samples maintained at 
         temperatures $T_\rS = 360\,{\rm K}$ and $T_\rD = 300\,{\rm K}$.
         By changing the flux $\Phi_\rG$ supplied to the gate the temperature $T_\rG$ 
         is changed so that the  transistor operates either as
         thermal switch or a thermal amplifier.\label{Flux} }.
\label{modes}
\end{figure}

The different operating modes of the transistor can be analyzed from the evolution of flux curves with respect to the gate temperature. These curves are plotted in Fig.~\ref{Flux} in the case of a silica source and
a silica drain with a VO$_2$ gate in between. We set $T_\rS=360\, {\rm K}$ and $T_D = 300\, {\rm K}$
and choose a separation distance between the source and the gate and between the gate and the drain 
to $d = 100\, {\rm nm}$. The thickness of the gate layer is set to $\delta = 50\, {\rm nm}$.

\begin{figure}
  \includegraphics[scale=0.35,angle=0]{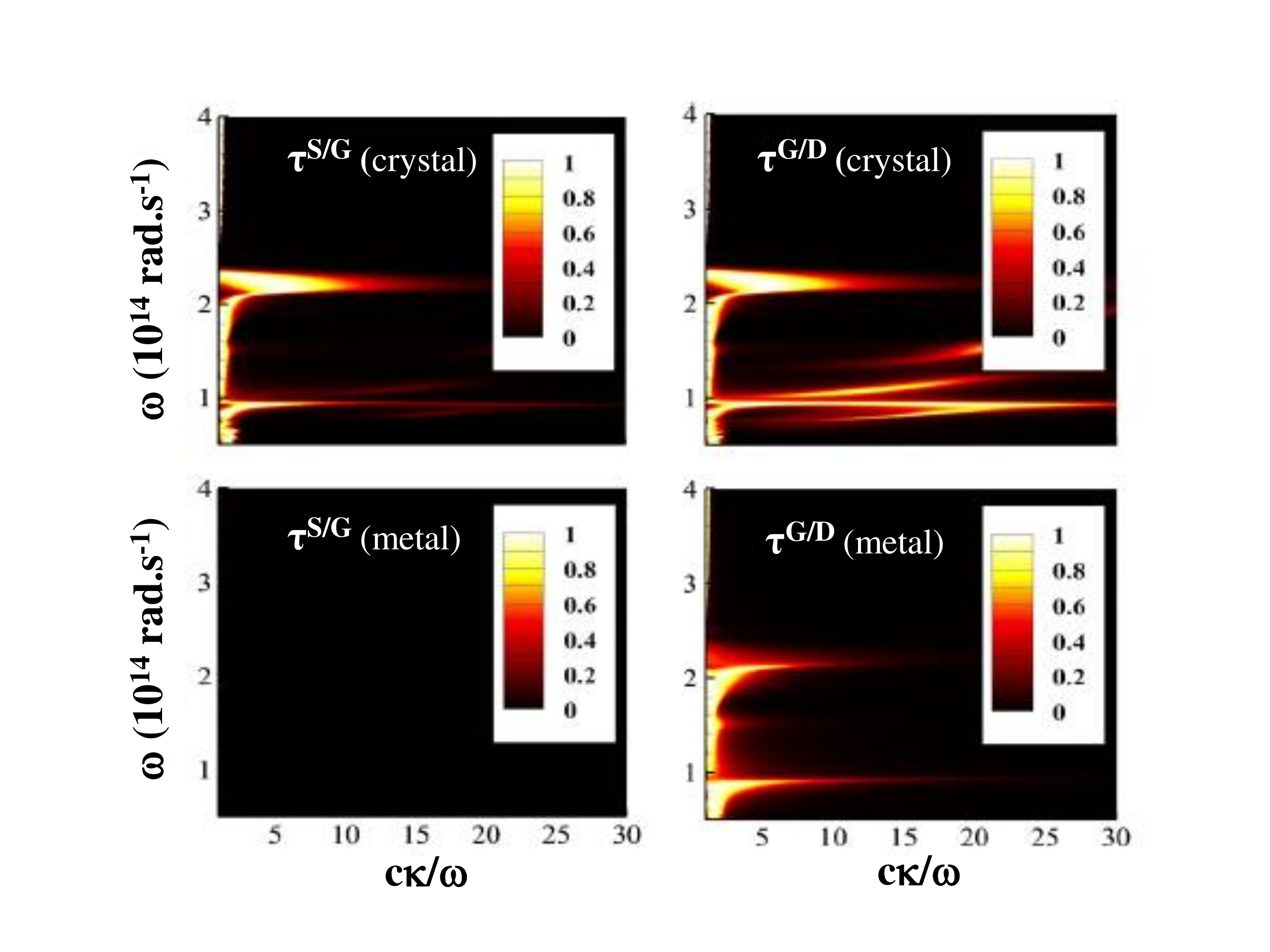} 
     \caption{Energy transmission coefficient of modes in the ($\omega,\boldsymbol{\kappa}$) plane for a SiO$_2$-VO$_2$-SiO$_2$ transistor with a $\delta=50\,{\rm nm}$ thick  gate and a separation distance of $d=100\,{\rm nm}$ between the source and the gate (resp. the drain and the gate).  
Wien's frequency (where the transfer is maximum) at $T=340\,{\rm K}$ is $\omega_{\rm Wien}\sim 1.3\times 10^{14}\,{\rm rad/s}$.}
  \label{transmission_coeff}
  \end{figure}

The equilibrium temperature of the gate,  obtained by solving the transcendental equation (\ref{Eq:equilibrium}) 
is for this configuration uniquely given by $T^{\rm eq}_G=332\,{\rm K}$ which is close to the critical temperature $T_\rc \approx 340\,{\rm K}$ of VO$_2$. Hence,in the steady-state situation, the gate is in its insulating phase. In this situation it supports surface phonon-polaritons~\cite{van Zwol1,PBA_PRL2014} in the mid-infrared range as well as the source and the drain so that these surface waves can couple together making the heat transfer very efficient between the source and the drain which can be seen by inspection of the transmission coefficients in Fig.\ref{transmission_coeff}. On the contrary, when the temperature of the gate is increased by external heating to values larger than $T_\rc$ then VO$_2$ undergoes a phase transition towards its metallic phase. In this case, the gate does not support surface wave resonances anymore so that the surface mode coupling between each solid element is suppressed (see Fig.\ref{transmission_coeff}) and the heat transfer drastically drops (Fig.~\ref{Flux}). This drastic change in the transfer of energy can be used to modulate the heat flux received by the drain by changing the gate temperature around its critical value. The thermal inertia of the gate as well as its phase transition delay defines the timescale at which the switch can operate. Usually the thermal inertia  limits the speed to some microseconds~\cite{Tschikin,Dyakov2} or even less~\cite{OrdonezEtAl2016}.

\begin{figure}
\includegraphics[scale=0.3,angle=0]{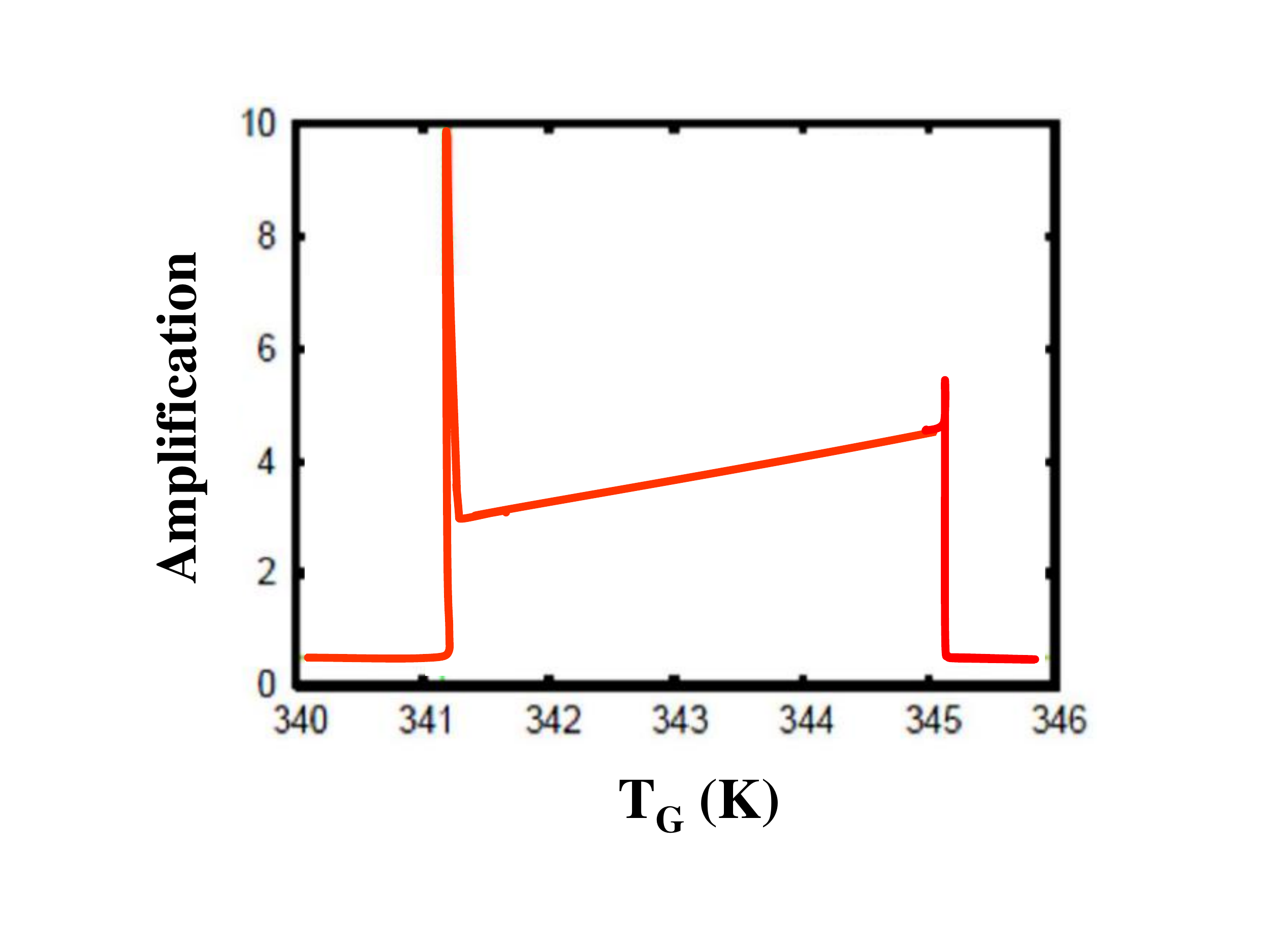} 
\caption{Amplification coefficient of a SiO$_2$-VO$_2$-SiO$_2$ transistor. When $T_G\ll T_c$ or $T_G\gg T_c$ the slopes of $\Phi_\rS$ and $\Phi_\rD$ are identical (modulo the sign) so that $\alpha=1/2$.  On the contrary, in the close neighborhood of $T_c$ we have $\alpha>1$ owing to the negative differential thermal resistance in the transition region. The large $\alpha$ at the beginning and the end of the phase-change region are artefacts due to the model used to describe the permittivity in this region. Here, the parameters of the transistor are the same as in Fig.~\ref{transmission_coeff}.}
\label{Amplification coefficient}
\end{figure}

But more interesting is the possibility offered by this system to amplify the heat flux received by the drain. This effect is the thermal analog of the classical transitor effect. To highlight this effect, let us focus our attention on the operating mode 
  in the region of phase transition around $T_\rc$. As we see in Fig.~\ref{modes} a small increase of $T_\rG$ leads to a drastic 
  reduction of flux received by the drain. As described by  Kats et al.\ in Ref.~\cite{Kats} this behavior can be associated  in far-field to a reduction of the thermal emission. This anomalous behavior corresponds to the so called negative differential thermal conductance as described  in~\cite{Fan}. The presence of a negative differential thermal conductance is a necessairy condition (but not sufficient) for observing a transistor effect. Indeed, the amplification coefficient of a transistor is defined as (see for example Ref.~\cite{Casati1})
  \begin{equation}
    \alpha \equiv \biggl|\frac{\partial \Phi_\rD}{\partial \Phi_\rG}\biggl| =  \frac{1}{\biggl|1 - \frac{\Phi_\rS'}{\Phi_\rD'}\biggr|}
  \end{equation}
  where
  \begin{equation}
    \Phi_{\rS/\rD}' \equiv \frac{\partial \Phi_{\rS/\rD}}{\partial T_\rG}.
  \end{equation}
By introducing the thermal resistances 
\begin{equation}
    R_D=\Phi_{\rD}^{'-1}
 \end{equation}
and
\begin{equation}
    R_S=-\Phi_{\rS}^{'-1}
 \end{equation}
 associated to the drain and to the source this coefficient can be recasted under the form
 \begin{equation}
    \alpha = \biggl|\frac{R_S}{R_S+R_D}\biggr|.
 \end{equation}
It immediately follows from this expression that when both resistances are positive, then $\alpha<1$. This, precisely happens outside the phase transition region (see Fig.~\ref{Flux})  where $\Phi_S' = - \Phi_D'$ so that $\alpha = 1/2$ 
(Fig.~\ref{Amplification coefficient}). On the contrary, in the region where the phase transition occurs this situation changes. A direct inspection of flux in this region (Fig.~\ref{Flux}) clearly shows that $\Phi_S'$ and $\Phi_D'$ have the same (negative) sign so that $R_S>0$ and $R_D<0$. It turns out that $\alpha > 1$ 
(Fig.~\ref{Amplification coefficient}).

%
%

\section{Radiative memory}

\begin{figure}
\includegraphics[scale=0.3]{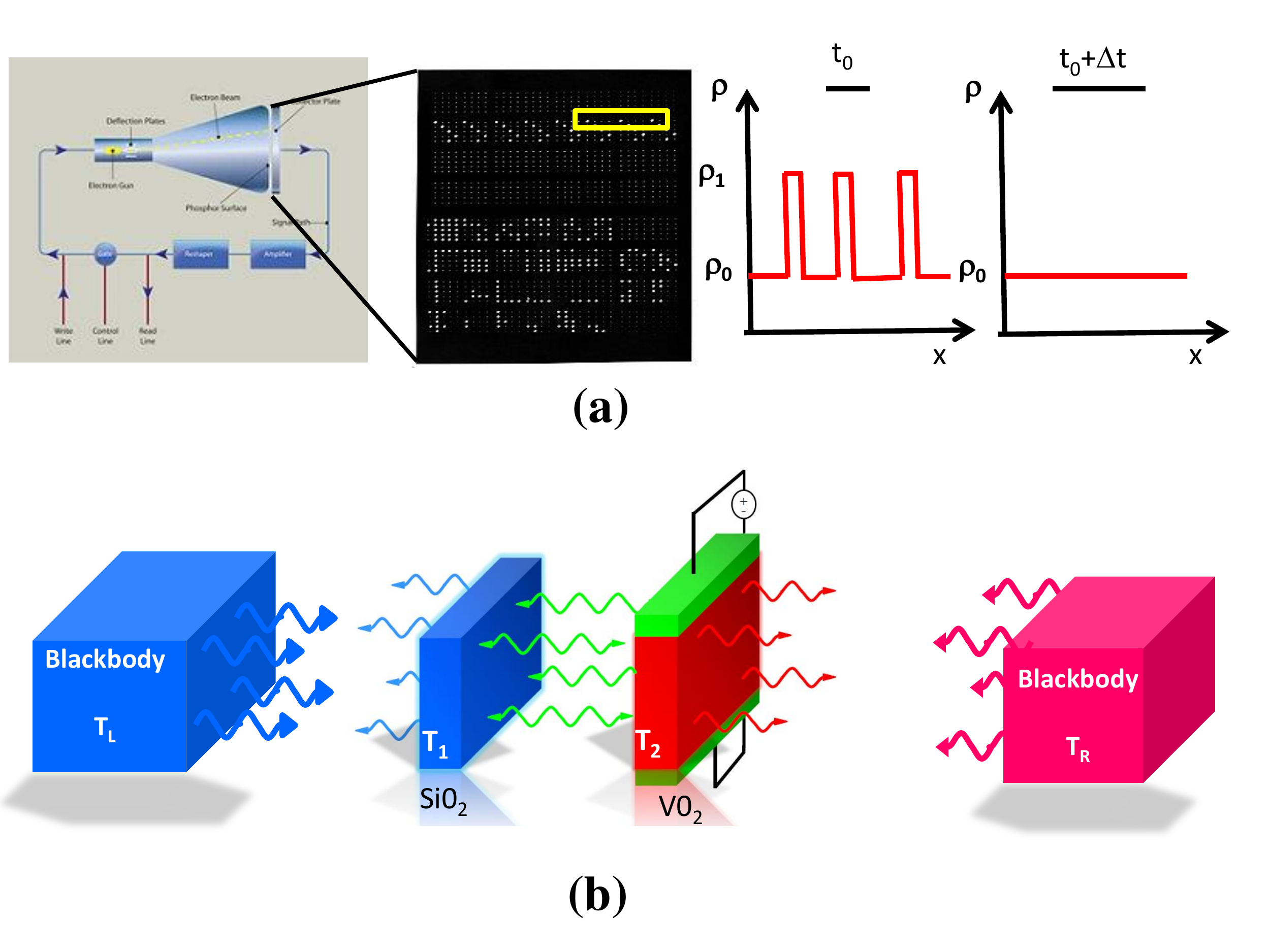}
\caption{(a) Sketch of an  electronic memory as proposed by Williams and Kilburn. It is based on writing/reading 
on a screen with an electron beam emitted by a cathode ray tube creating a local distribution of charges during a finite time interval $\Delta t$.  By scanning the screen  it is possible to store multiple data bits and also to read them.(b) Radiative thermal memory: a membrane made of an IMT material is placed at a distance 
$d$ from a dielectric layer. The system is surrounded by two thermal baths at different temperatures $T_{\rm L}$ and $T_{\rm R}$. 
The temperature $T_2$ can be increased or reduced either by Joule heating by applying a voltage difference through a couple of 
electrodes or by using Peltier elements.} 
\label{memory}
\end{figure}

In 1946  Williams and  Kilburn developed a memory to store data electronically using a cathode ray tube. The basic idea of their invention is summarized in Fig.~\ref{memory}(a). A small electrostatic charge appears on the surface of a screen  lighted up by an electron beam and this charge remains for a short period of time before leaking away. This invention constitutes the first volatile memory. Information is stored as long as the charge is not dissipated.
In this section, we describe the possibility to store information or energy for arbitrary long time using thermal photons~\cite{Slava}. The concept of thermal memory is closely related to the thermal bistability of a system, i.e.\ the presence of at least two equilibrium temperatures~\cite{BaowenLi3}. 

To illustrate this, let us consider the system as depicted in Fig.~\ref{memory}(b) composed by two 
parallel homogeneous membranes made of VO$_2$ and SiO$_2$. These slabs have finite thicknesses $\delta_{1}$ and $\delta_{2}$ 
and are separated by a vacuum gap of distance $d$. The left (right) membrane is illuminated by the field radiated by a blackbody 
of temperature $T_{\rm L}$ ($T_{\rm R}$), where $T_{\rm L} \neq T_{\rm R}$. The membranes themselves 
interact on the one hand through the intracavity fields and on the other with the field radiated by the two black bodies.

The heat flux across any plane $z = \bar{z}$ parallel to the interacting surfaces can  be evaluated
using Rytov's fluctuational electrodynamics~\cite{Rytov,Polder}. Here, we consider the situation where
the separation distance $d$ is large enough compared to the thermal wavelengths 
[i.e. $d\gg\max(\lambda_{T_i}=c\hbar/(\kb T_{i})$, $i = 1,2,{\rm L}, {\rm R}$ ] so that near-field heat exchanges 
can be neglected (for a thermal memory working in near-field regime see Ref.~\cite{DyakovMemory}). 
In this case we obtain
\begin{equation}
\begin{split}
     \varphi(\bar{z}) = 2\epsilon_0c^2 \!\!\sum \limits_{\underset{\phi=\left\{ +,-\right\}}{j=s,p}}\int_{0}^{\infty}\!\!\frac{d\omega}{2\pi}\int\!\!\! \frac{{\rm d}^2 \boldsymbol{\kappa}}{(2 \pi)^2}\frac{\phi k_{z0}}{\omega} \mathfrak{C}_j^{\phi,\phi}(\omega,\boldsymbol\kappa),
\label{Eq:base}
\end{split}
\end{equation}
where the field correlators
\begin{equation}
\mathfrak{C}_j^{\phi,\phi'}(\omega,\boldsymbol{\kappa})= \frac{1}{2} \langle[E_j^\phi(\omega,\boldsymbol{\kappa})E_j^{\phi'\dagger}(\omega,\boldsymbol{\kappa})+E_j^{\phi'\dagger}(\omega,\boldsymbol{\kappa})E_j^\phi(\omega,\boldsymbol{\kappa})]\rangle
\end{equation}
of local field amplitudes in polarization $j$ can be expressed~\cite{Slava} in terms of 
reflection and transmission operators $\mathfrak{R}_i^{\pm}$ and $\mathfrak{T}_i^{\pm}$  of the layer $i$  toward the right ($+$) and the left ($-$), as
\begin{equation}
\begin{split}
  \mathfrak{C}_1^{+,+}=\mathcal{S}(T_1)(1- \mid \mathfrak{R}_1^{+} \mid^2 -\mid \mathfrak{T}_1^{+} \mid^2 ),\\
  \mathfrak{C}_2^{-,-}=\mathcal{S}(T_2)(1- \mid \mathfrak{R}_2^{-} \mid^2 -\mid \mathfrak{T}_2^{-} \mid^2 ),\\
  \mathfrak{C}_L^{+,+}=\mathcal{S}(T_L),\;\;\;\;\;\;\;\;\;\;\;\;\;\;\;\;\;\;\;\;\;\;\;\;\;\;\;\;\;\;\;\;\;\;\\
  \mathfrak{C}_R^{-,-}=\mathcal{S}(T_R),\;\;\;\;\;\;\;\;\;\;\;\;\;\;\;\;\;\;\;\;\;\;\;\;\;\;\;\;\;\;\;\;\;\;\\
\end{split}\label{Eq:cavity5}
\end{equation}
with 
\begin{equation}
  \mathcal{S}(T)=\pi\frac{\omega}{\epsilon_0 c^2}\Theta(\omega,T)\Re\biggl(\frac{1}{k_z}\biggr).
\end{equation}
Here $k_z$ denotes the normal component of wave vector in the medium of consideration. Using expression (\ref{Eq:base}) we can calculate the
net flux $\Phi_1=\varphi(0)-\varphi(-\delta_1)$ [ $\Phi_2=\varphi(d+\delta_2)-\varphi(d)$] received by the first (second) membrane.

\begin{figure}
\includegraphics[scale=0.3]{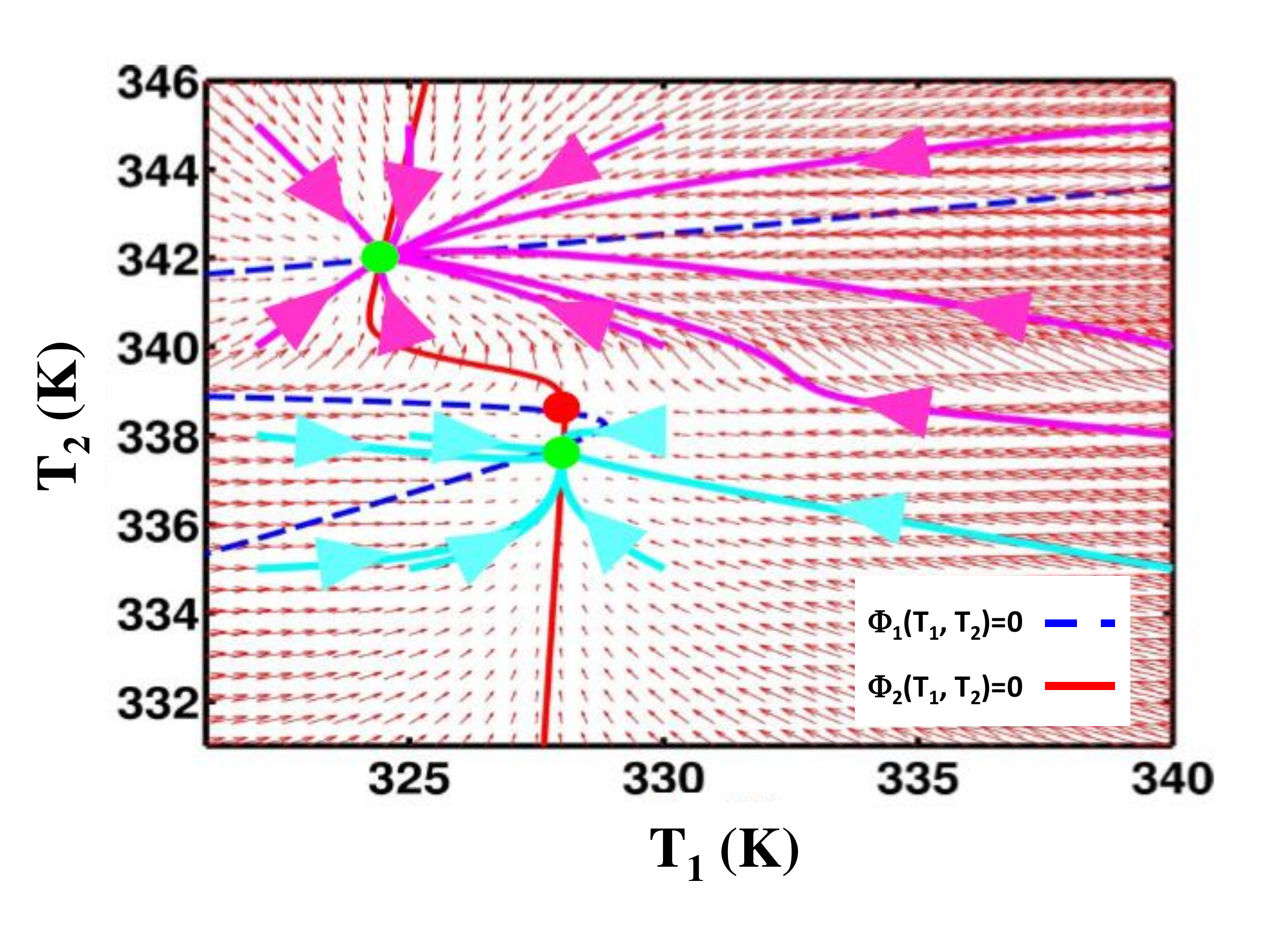}
\caption{Phase portrait of a two-membrane SiO$_2$/VO$_2$ system of thickness $\delta_1=\delta_2=1\,\mu{\rm m}$. The  blue dashed and red solid lines represent the local equilibrium conditions $\Phi_1 = 0$ and $\Phi_2 = 0$ of each membrane. The green (red) points denote the unstable (stable) global steady-state temperatures, $(T_1^{(2)},T_2^{(2)}) = (328.06\,{\rm K},338.51\,{\rm K})$, $(T_1^{(1)},T_2^{(1)}) = (328.03\,{\rm K},337.77\,{\rm K})$, , and  $(T_1^{(3)},T_2^{(3)}) = (324.45\,{\rm K},341.97\,{\rm K})$. The red arrows represents the vector field $\boldsymbol{\Phi}$. The temperature of thermal reservoirs (black bodies) are $T_{\rm L}=320\,{\rm K}$ and $T_{\rm R}=358\,{\rm K}$.} 
\label{bistability}
\end{figure} 

The phase portrait of two membranes plotted in Fig.~\ref{bistability} shows all the time evolution of  SiO$_2$ and VO$_2$ membranes for all initial conditions.  In this figure, 
the dashed blue (solid red) line represents the local equilibrium temperatures for the first (second) membrane 
that is the set of temperatures couples ($T_1,T_2$) which satisfy the condition $\Phi_1(T_1,T_2)=0$ [$\Phi_2(T_1,T_2)=0$]. 
The intersection of these two lines define the global steady-state temperatures of the system. It clearly appears that the system possesses three equilibrium temperatures. However, as shown in ~\cite{Slava} with the help of a stability study of the system,
only two of three equilibrium points $\mathbf{T}\equiv (T^{eq}_1,T^{eq}_2)^t$ that appear in 
Fig.~\ref{bistability} are stable.

\begin{figure}
\includegraphics[scale=0.3,angle=-90]{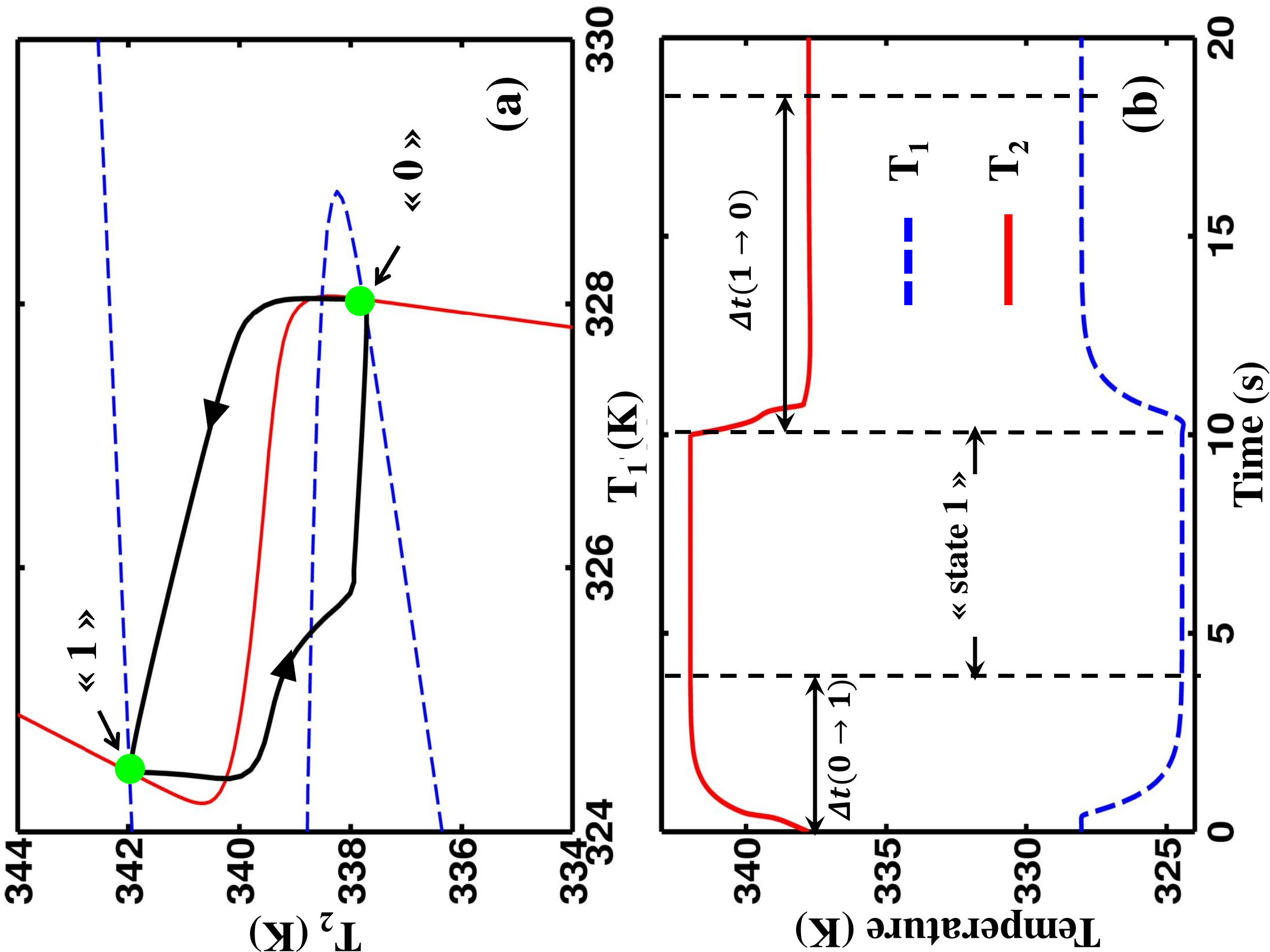}
\caption{(a) Time evolution of a thermal memory made of a SiO$_2$/VO$_2$ system 
with $\delta_1=\delta_2=1\,\mu{\rm m}$ during a transition between the thermal states "0" and  "1" . 
The volumic powers supplied and extracted from the VO$_2$ layer during a time interval $\Delta t_1=0.4\,{\rm s}$ and $\Delta t_2=1.5\,{\rm s}$  
are $Q_2=10^{-2}\,{\rm W} {\rm mm}^{-3}$ and $Q_2=-2.5\times10^{-2}{\rm W}{\rm mm}^{-3}$, respectively. The writing time of state "1" ("0") from the state "0" ("1") is  
$\Delta t(0\rightarrow 1)=4\,{\rm s}$ ($\Delta t(1\rightarrow 0)=8\,{\rm s}$). (b) Time evolution of temperature $T_1(t)$ and $T_2(t)$ of SiO$_2$ and VO$_2$ membranes. 
The thermal states "0" and "1" can be maintained for arbitrary long time provided that the thermostats ($T_L=320\,{\rm K}$ and $T_R=358\,{\rm K}$) remain switched on.} 
\label{memory2}
\end{figure}

The two stable thermal states can naturally be identified to the "0" and  "1" of one bit of information. As long as the temperatures of the two reservoirs are held constant, the system remains in the same stable thermal state. By perturbating the system (for instance by adding or extracting a certain power to one membrane) it is possible to switch from one thermal state to the other state as shown in Fig.~\ref{memory2}. The temporal dynamics of temperatures $T_1$ and $T_2$ of the two membranes are solution of the energy balance equation
\begin{equation}
  \partial_t \mathbf{T} = \boldsymbol{\Phi} + \mathbf{Q}
  \label{Eq:diff}
\end{equation}
where we have introduced the vectors $\mathbf{T}=(T_1(t),T_2(t))^t$,
 $\boldsymbol{\Phi} \equiv \bigl(\Phi_1(T_1,T_2)/I_1, \Phi_2(T_1,T_2)/I_2\bigr)^t$,
and $\mathbf{Q} \equiv (Q_1\delta_1/I_1,Q_2\delta_2/I_2)^t$. In this nonlinear system of coupled differential equations  $Q_i$ ($i = 1,2$) is the power per 
unit volume which can be added to or extracted from both membranes by applying a voltage difference 
through a couple of electrodes as illustrated in Fig.~\ref{memory} or by using Peltier elements.
In the $\mathbf{Q}$ vector, the terms  $I_i \equiv C_i \rho_i \delta_i$ (i=1,2) represent the thermal inertia of both membranes 
where $C_i$ and $\rho_i$ denote the heat capacity and the mass density of each material [see Suppl.\ Mat.\ of Ref.~\cite{Slava} for more details].

The resolution of this system from the two different initial conditions corresponding to the equilibrium states allows us to describe the dynamics of the switching process between one state and the other one. 
Here below we briefly describe this writing-reading procedure\cite{Slava}. To this end, we consider the SiO$_2$-VO$_2$ 
system made with membranes of equal thicknesses $\delta_1 = \delta_2=1\,\mu{\rm m}$ which are
coupled to two reservoirs of temperatures $T_\rL=320\,{\rm K}$ and $T_\rR=358\,{\rm K}$. 
Let us define "0" as the thermal state at the temperature $T_2=\min(T^{(1)}_2,T^{(3)}_2)$. 
To make the transition towards the thermal state "1" the VO$_2$ membrane must be heated. 

Step 1 (transition from the state "0" to the state "1"): A volumic power $Q_2=10^{-2}\,{\rm W}{\rm. mm}^{-3}$ 
is added to this membrane during a time interval $\Delta t_1\thickapprox 0.4\,{\rm s}$ to reach a region in the plane ($T_1,T_2$) [see  Fig.~\ref{memory2}(a)]
where all trajectories converge naturally (i.e.\ for $Q_2=0$) after some time toward the state "1", the overall transition time 
is $\Delta t(0\rightarrow 1)=4 s$  [see Fig.~\ref{memory2}(b)].

Step 2 (maintaining the stored thermal information): Since the state "1" is a fixed point, the thermal 
data can be maintained for arbitrary long time provided that the thermal reservoirs are switched on. 

Step 3 (transition from the state "1" to the state "0"): Finally, a volumic power $Q_2=-2.5\times 10^{-2}\,{\rm W}{\rm. mm}^{-3}$ 
is extracted from the VO$_2$ membrane during a time interval $\Delta t_2\thickapprox 1.5\,{\rm s}$ to reach a region 
[below $T_2=338\,{\rm K}$ in Fig.~\ref{memory2}(a)] of natural convergence to the state "0" . In this case the transition 
time becomes $\Delta t(1\rightarrow 0)=8\,{\rm s}$. Compared with its heating, the cooling of VO$_2$ does not follow the 
same trajectory [see Fig.~\ref{memory2}(a)] outlining the hysteresis of the system which accompanies its bistable behavior. 
To read out the thermal state of the system a classical electronic thermometer based on the thermal 
dependence of the electric resistivity of membranes can be used.

These results show that  many-body radiative systems could be used as volatile thermal memory. The radiative bistability which exists in  some of these media can  be exploited for energy storage both at macroscale (far-field regime) and subwavelength scale (near-field regime). This thermal energy could in principle
release heat upon request in its environment making these systems active building blocks for a smart  management of heat exchanges between different objects without any contact.

%
%

\section{Logic gates with thermal photons}

The next-generation of internet of things infrastructures has the potential to change the way people and systems live in a world of massive and disparate data sources, and to provide opportunities for connectivity at different scales.  Instead of using electrical signals, purely thermal signals could be used. However the development of such a technology requires the existence of thermal logic gates being able to perform a boolean information treatment as their electronic counterpart do. In this section we demonstrate that  NOT, OR and AND gates can be realized exploiting the radiative heat
exchanges in N-body systems with phase-change materials~\cite{PBA_PRL_logic}.
To start, let us consider the simplest logic gate, the NOT gate which implements a logical negation. This operation can be performed by using the simple termal transistor discussed before [see Fig.~\ref{transistor}(b)].  In order to operate as a NOT gate the
temperature of the source is maintained at a fixed temperature $T_{\rm S}=360\,{\rm K}$. 

\begin{figure}
\includegraphics[scale=0.35]{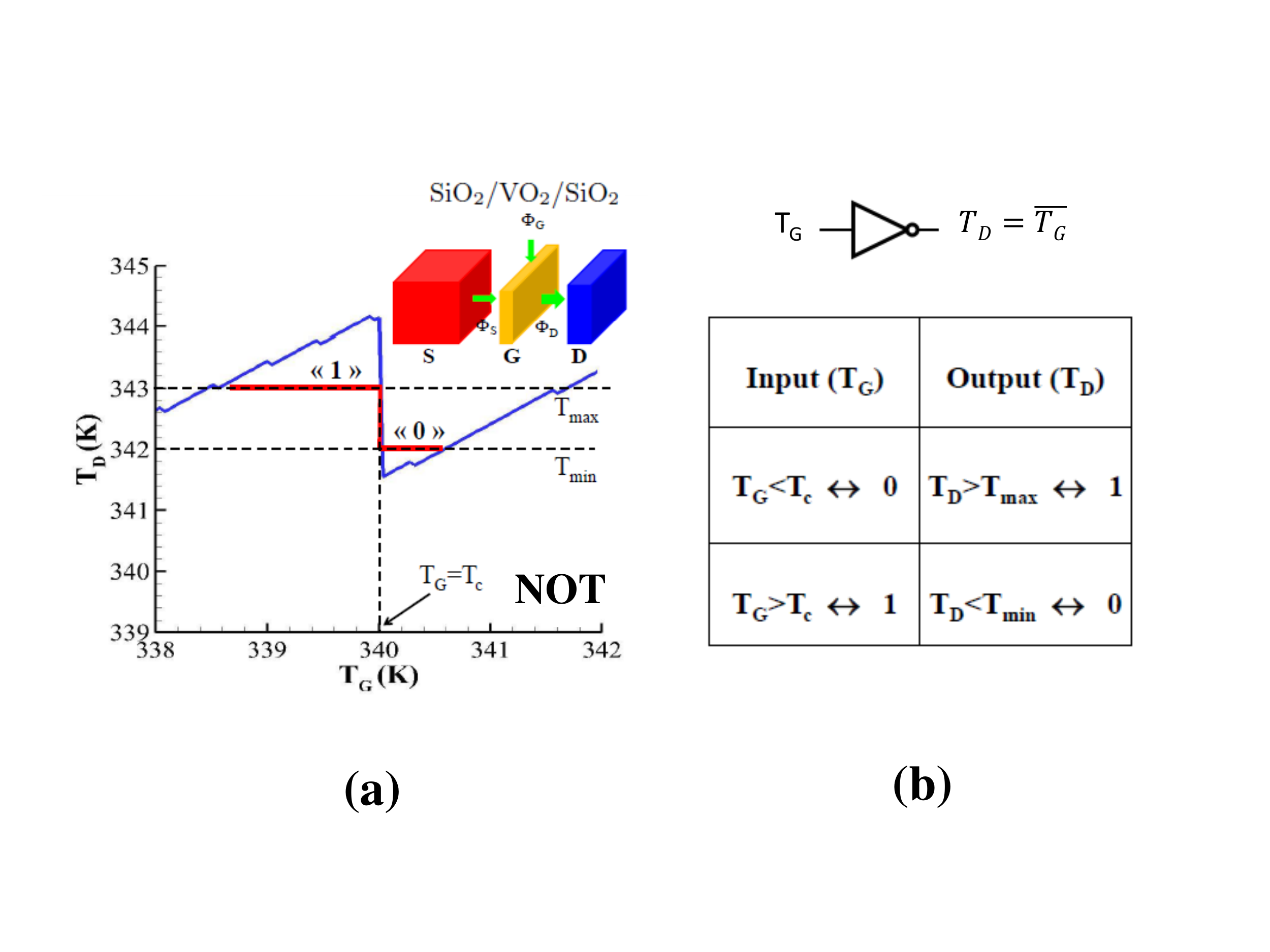}
\caption{(a) NOT gate made with a SiO$_2$/VO$_2$/SiO$_2$ plane thermal transistor. The source (S) is assumed to be semi-infinite while the gate (G) and the drain (D) have a tkickness of 100 nm. The separation  distances are d=100 nm. The temperature $T_G$ of the gate defines the input and the drain temperature sets the gate output. In the operating range of the gate around the critical temperature $T_c$ of the gate, if $T_D<T_{min}$ the gate is in the thermal state "0". On the contrary,  when $T_D>T_{max}$ the gate is in the state "1". The rectangular function represents the ideal response of the gate in its operating range.(b) Truth table for the NOT gate.} 
\label{NOT_gate}
\end{figure}

Then, as shown in Fig.~\ref{NOT_gate},  the temperature of the intermediate layer plays the same role as an input while the drain temperature acts as an output and defining the thermal state with $T_{\rm G} < T_{\rm c}$ as the '0' state and the 
termal state with  $T_{\rm G} > T_{\rm c}$ as the '1' state, then the system works as a NOT gate. The input state can be controlled from outside by adding or removing heat from the intermediate gate layer. 

\begin{figure}
\includegraphics[scale=0.35]{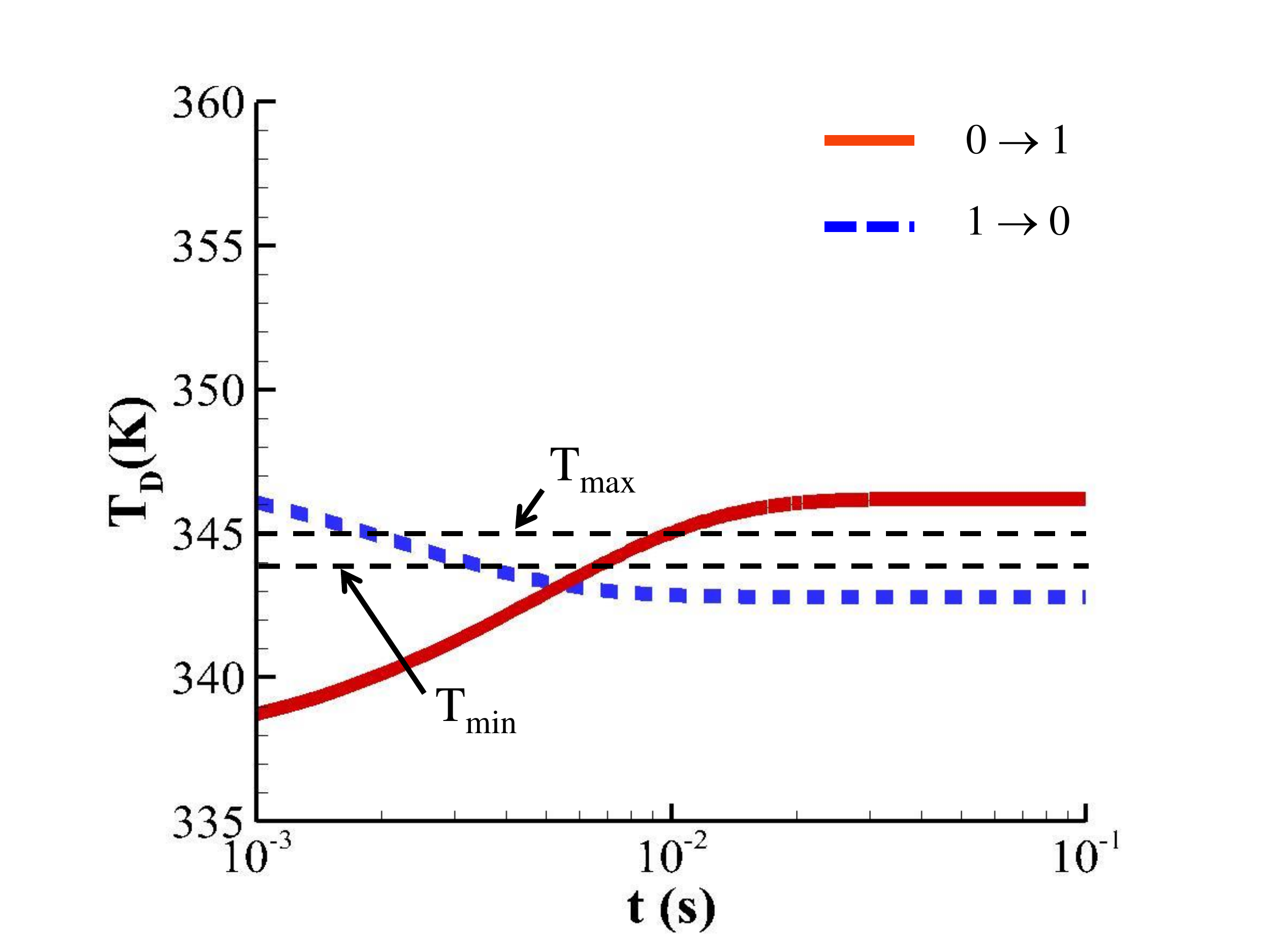}
\caption{Transition dynamics of  a SiO$_2$/VO$_2$/SiO$_2$ NOT gate with the same geometrical parameters as in Fig.~\ref{NOT_gate}. The red solid (blue dashed)  line shows the time evolution of the drain temperature from the state "0" ("0") to the state  "1" ("0")  when the initial temperature is $T_D(0)=337 K$ ($T_D(0)=348 K$ ). The gate temperature is assumed to be fixed at $T_G=343 K$ ($T_G=337 K$ ) and the threshold temperatures are $T_{min}=344 K$ and  $T_{max}=345 K$.} 
\label{dyn NOT}
\end{figure}

The dynamics of a NOT gate for passing from one state to another (given two different values for $T_D(t=0s)$ corresponding to '0' and '1')  can be 
evaluated  by solving the nonlinear dynamical equation
\begin{equation}
\rho C d_\rD \partial_t T_D=\Phi_\rD(T_S,T_G,T_D),
\label{dynamic NOT gate}
\end{equation}
where $\Phi_\rD(T_S,T_G,T_D)$ is the flux received by the drain while $\rho$, $C$ and $d_\rD$  denote the mass density, the heat capacity and the thickness of the drain, respectively. In Fig.~\ref{dyn NOT} we show the 
transition dynamics from the state '0' to the state '1'  (from the state '1' to the state '0') with an initial temperature $T_\rD(0)=337 K$ 
($T_\rD(0)=348 K$ ) and a gate at fixed temperature $T_\rG=343 K$ ($T_\rG=337 K$). The overall time the NOT gate takes to switch from state '0' to state '1' (state '1' to state '0') is on the order of a few ms. Note that compared with the operating speed of electric logic gates this time is  large because of the thermal inertia. Nevertheless to activate thermal sensors this delay is largely sufficient.

\begin{figure}
\includegraphics[scale=0.35]{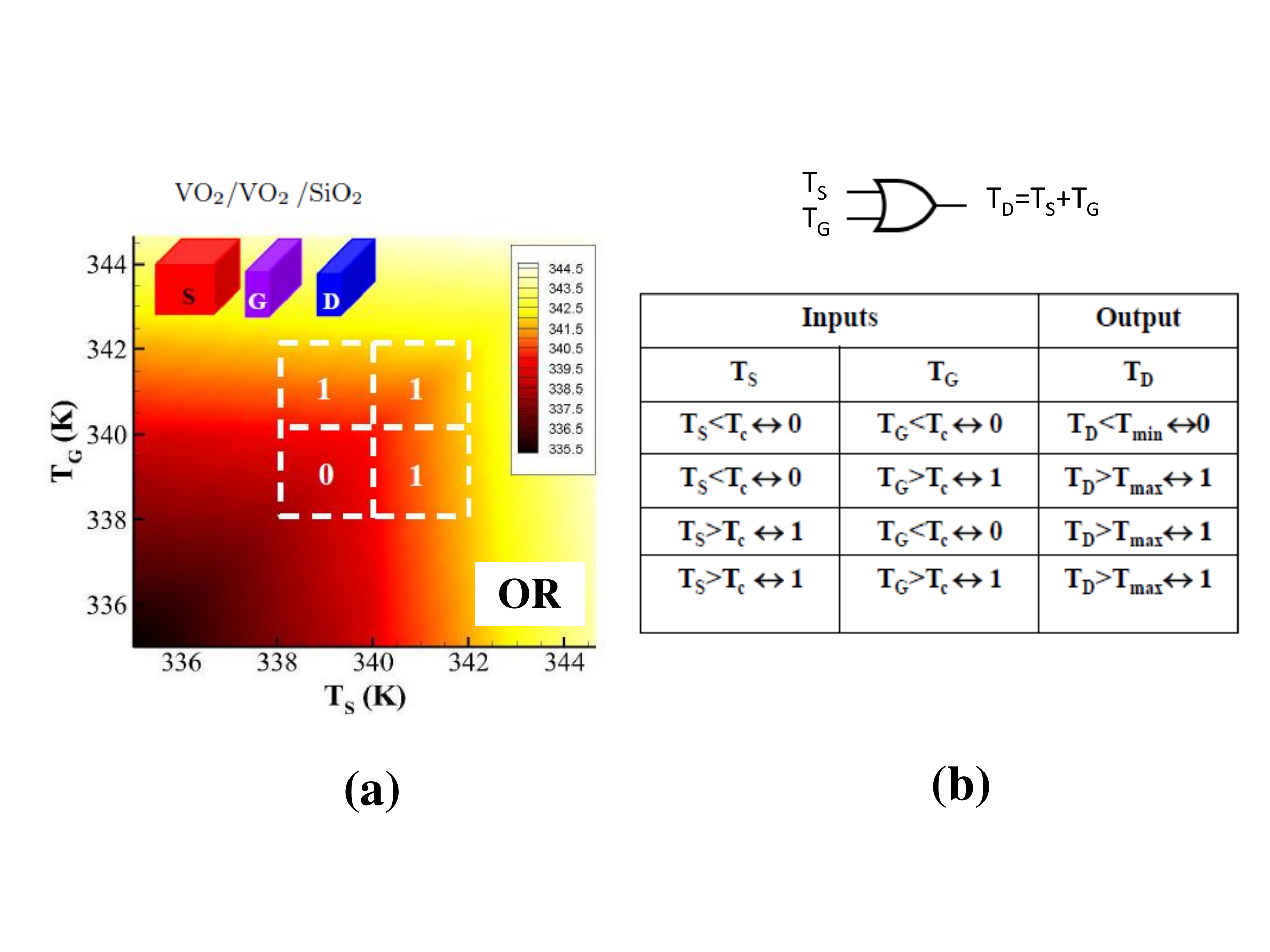}
\caption{(a) OR gate made with a VO$_2$/VO$_2$ /SiO$_2$ thermal transistor. The source (S) is assumed to be semi-infinite while the  gate (G) and the drain (D) have a tkickness of 100 nm. The separation  distances are d=100 nm. The two temperatures $T_{S}$ and  $T_{G}$ define the inputs and the drain temperature sets the gate output. The operating range of the gate is limited to the dashed rectangular domain centered at $(T_{S},T_{G})=(T_c,T_c)$  with $T_c=340 K$ .  (b) Truth table for the OR gate.} 
\label{OR_gate}
\end{figure}

Now let us show that a many-body system can also operate as an OR gate.  Note that we consider here, contrary to the NOT gate, a transistor with a phase-change material for both the source and the gate layer while silica is used for the drain layer. The temperatures of the source and the gate are used as inputs while the drain temperature sets the output of the logic gate. The temperature evolution of the drain with respect to the temperatures of the source and the gate is plotted in Fig.~\ref{OR_gate}.  We see that around the point  $(T_{\rm S},T_{\rm G})=(T_\rc,T_\rc)$  with $T_\rc=340 K$ where the phase change occurs both in the source and the gate, the temperature of the drain undergoes a significant variation.   If $T_{\rm S}$ and $T_{\rm G}$ are both smaller than the critical temperature $T_\rc$ of VO$_2$ then these two elementary blocks behave like a dielectric so that the temperature of the drain is small. On the contrary, if either the source or the gate undergo a phase change, then the temperature of the drain increases abruptly. Hence, by conveniently intoducing two suitable threshold temperatures $T_{\rm min}$ and $T_{\rm max}$ it is clear that we can associate to the drain two different thermal states with respect to the temperatures $T_\rS$ and $T_\rG$  around the region $(T_{\rm S},T_{\rm G})=(T_\rc,T_\rc)$. Therefore, this transistor behaves like an OR gate with the truth table given in Fig.~\ref{OR_gate}(b). Note also that by reversing the definition of thermal states in the drain, this system mimicks a NOR gate.

\begin{figure}
\includegraphics[scale=0.35]{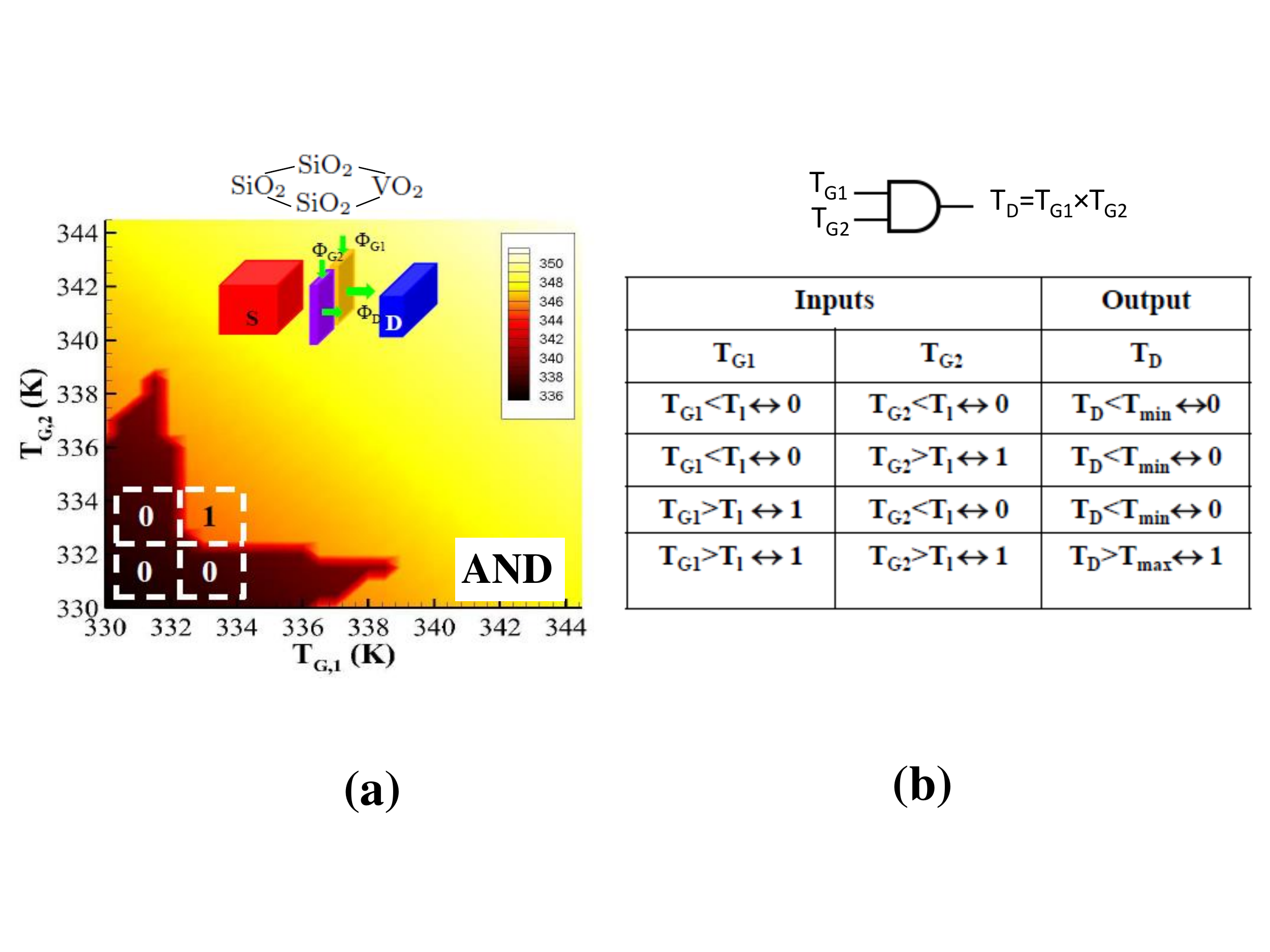}
\caption{(a) AND gate made with a double SiO$_2$ gate thermal transistor, the source is made in SiO$_2$ while the drain is in VO$_2$. The source (S) is assumed to be semi-infinite while both gates (G1, G2)  and the drain (D) have a tkickness of 250 nm and 500 nm respectively. The separation  distances are d=100 nm and the gates are assumed isolated one from the other. The two temperatures $T_{G1}$ and  $T_{G2}$ define the inputs and the drain temperature sets the gate output. The operating range of the AND gate is limited to the dashed rectangular domain centered at $(T_{G1},T_{G2})=(T_l,T_l)$  with $T_l=332.2 K$ . (b) Truth table for the ideal AND gate. } 
\label{AND_gate}
\end{figure}

Finally, we sketch the realization of an AND gate. This double input device is shown in Fig.~\ref{AND_gate}(a). It is a double gate thermal transistor made with two silica gates and a silica source. Contrary to the NOT gate it is the drain which is made of a phase-change material. The temperatures of both gates set the two inputs of the logic gate. 
We  assume that the temperatures of the two gates can now be controlled independently making the assumption, for convenience, that they are thermally insulated one from the other so that we can express the heat flux received by the drain just as the mean value of two NOT gates, i.e.\ we have
\begin{equation}
\begin{split}
  \phi_{\rD} &= \frac{1}{2}\sum_{j = \{\rm s,p\}}\int\! \frac{{\rm d}^2 \boldsymbol{\kappa}}{(2 \pi)^2} \, \bigl[\Theta_{\rS,\rG1}(\omega)\mathcal{T}^{\rS/\rG1}_j(\omega,\boldsymbol{\kappa}; d)\\
                &\quad+\Theta_{\rG1,\rD}(\omega)\mathcal{T}^{\rG1/\rD}_j(\omega,\boldsymbol{\kappa}; d)+\Theta_{\rS,\rG2}(\omega)\mathcal{T}^{\rS/\rG2}_j(\omega,\boldsymbol{\kappa}; d)\\
&\quad+\Theta_{\rG2,\rD}(\omega)\mathcal{T}^{\rG2/\rD}_j(\omega,\boldsymbol{\kappa}; d)\bigr].
\end{split}
\label{Eq:Flux_D_2}
\end{equation}
Here the transmission coefficients are given by the same expressions as in (\ref{Trans1}) for a single gate transistor. In Fig.~\ref{AND_gate}(a) we show the equilibrium temperature $T_\rD$ of the drain with respect to the gate temperatures. In the central region around  $(T_{G1},T_{G2})=(T_l,T_l)$ with $T_l=332.2 K$ we see that $T_\rD$ can undergo a sudden variation after a short change in the gate temperatures. By introducing two critical temperatures $T_{min}$ and  $T_{max}$ we can associate to the drain two thermal states '0' or '1'  with respect to the relative value of the drain temperature to these thresholds. As it clearly appears in Fig~\ref{AND_gate}(a). this double gate system behaves as a digital AND gate with the truth table given in Fig.~\ref{AND_gate}(b). Note that by reversing the definition of thermal states '0' and '1' the AND gate works as a NAND gate. To go beyond this elementary logical operation a combination of various logical gates will now be necessary. However, in near-field regime this combination cannot be sequential, because of many-body effects which make the heat transport throughout the structure non additive.This demands for the development of a general many-body theory.

%
%

\section{magnetic control of heat flux}

In this last section we discuss the last developments on the tunability of radiative heat transfer in N-body systems. Although the radiative heat transport along nanoparticle systems have been intensively studied~\cite{Brongersma,PBAAPL2006,PBAPRB2008,Ordonnez2016} since two decades, the external control of radiative heat transfers in complex networks remains largely an open problem. Here below we discuss the possibility, recently demonstrated~\cite{PBAHall}, of a strong tunability of heat transfers in magneto-optical networks with an external magnetic field. This magnetic control of heat flux is associated to the presence of a photon thermal Hall effect in these systems. 

The classical Hall effect~\cite{Hall}  discovered by Edwin Hall at the end of the 19th century results in the appearance of a transverse electric current inside a conductor under the action of an external magnetic field applied in the direction orthogonal to the primary voltage gradient. This effect comes from the Lorentz force which acts transversally on the electric charges in motion through the magnetic field curving so their trajectories. Very shortly after this discovery, a thermal analog of this effect has been observed by Righi and Leduc~\cite{Leduc} when a temperature gradient is applied throughout an electric conductor. As for the classical Hall effect, this effect is intrinsically related to the presence of free electric charges. So, one can not expect a thermal Hall effect with neutral particles. Nevertheless, during the last decade researchers have highlighted such an effect in non-conducting materials due to phonons~\cite{Strohm,Jin} or magnons (spin waves)~\cite{Hirschberger,Fujimoto,Katsura,Onose}. 

\begin{figure}
\includegraphics[scale=0.3]{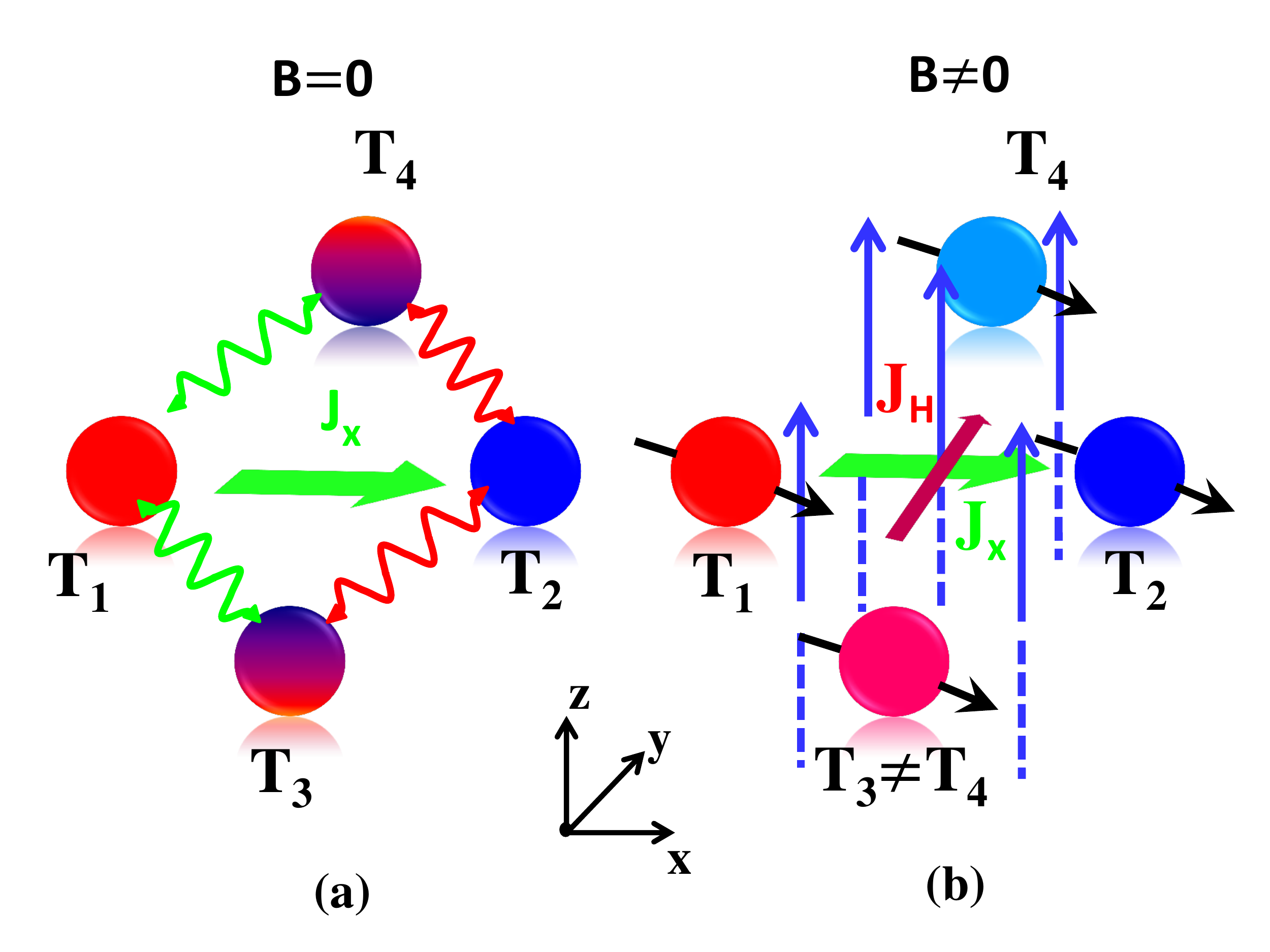}
\caption{Sketch of the four terminal junction made with magneto-optical nanoparticles. (a) Without magnetic field ($B=0$) the particles are optically isotropic so that the system is thermally symmetric (i.e. $T_3=T_4$) and the Hall flux $\varphi_H\equiv \varphi_y$ is zeroe. (b) If a magnetic field is applied in the $z$ direction, the particles become anisotropic breaking the apparent symmetry of system so that a temperature gradient is generated in the $y$ direction giving rise to a non-zero Hall flux. The black arrows crossing the particles illustrate their anisotropy.} 
\label{Hall effect}
\end{figure}

Here below we investigate the near-field heat exchanges in a four-terminal  system (see Fig.~\ref{Hall effect}) which is composed of magneto-optical particles under the action of a constant magnetic field applied perpendicularly to the system. Those particles can exchange electromagnetic energy between them. By connecting the two particles along the $\mathbf{x}$-axis to two heat baths at two different temperatures, a heat flux flows through the system between these two particles. Without external magnetic field (Fig.~\ref{Hall effect}(a)) all particles are isotropic, so that the two others unthermostated particles have, for symmetry reasons, the same equilibrium temperatures and therefore they do not exchange heat flux through the network. On the contrary, when a magnetic field is applied orthogonally to the particle network (Fig.~\ref{Hall effect}(b)), the particles become anisotropic so that the symmetry of the system is broken (Fig.~1). As we will see hereafter, when the steady-state regime is reached, the two unthermostated particles display two different temperatures. Therefore a heat flux propagates transversally to the primary applied temperature gradient.

Using the Landauer formalism for N-body systems~\cite{PBAEtAl2011,Riccardo,PRL_superdiff,Nikbakht,Incardone} the heat flux exchanged between the $i^{th}$ and the $j^{th}$ particle in the network reads 
\begin{equation}
  \varphi_{ij}=\int_{0}^{\infty}\frac{\rd\omega}{2\pi}\,[\Theta(\omega,T_{i})\mathcal{T}_{i,j}(\omega)-\Theta(\omega,T_{j})\mathcal{T}_{j,i}(\omega)]\label{Eq:InterpartHeatFlux},
\end{equation}
where $\mathcal{T}_{i,j}(\omega)$ denotes the transmission coefficient, at the frequency $\omega$, between the two particles. When the particles are small enough 
compared with their thermal wavelength $\lambda_{T_{i}} = c\hbar/(\kb T_{i})$ ($c$ is the vacuum light
velocity, $2 \pi \hbar$ is Planck's constant, and $\kb$ is Boltzmann's constant) they can be modeled by simple radiating electrical dipoles.
In this case the transmission coefficient is defined as~\cite{Nikbakht}
\begin{equation}
  \mathcal{T}_{i,j}(\omega)=2\Im\Tr\bigl[\mathds{A}_{ij}\Im\bar{\bar{\boldsymbol{\chi}}}_j\mathds{C}_{ij}^{\dagger}\bigr],
\end{equation}
where $\bar{\bar{\boldsymbol{\chi}}}_j$, $\mathds{A}_{ij}$ and $\mathds{C}_{ij}$ are the susceptibility tensor plus two 
matrices which read~\cite{Nikbakht} in terms of free space Green tensor $\bar{\bar{\boldsymbol{G}}}_{ij}^{0}=\frac{\exp({\rm i}kr_{ij})}{4\pi r_{ij}}\left[\left(1+\frac{{\rm i}kr_{ij}-1}{k^{2}r_{ij}^{2}}\right)\mathds{1}+\frac{3-3{\rm i}kr_{ij}-k^{2}r_{ij}^{2}}{k^{2}r_{ij}^{2}}\widehat{\mathbf{r}}_{ij}\otimes\widehat{\mathbf{r}}_{ij}\right]$
($\widehat{\mathbf{r}}_{ij}\equiv\mathbf{r_{\mathit{ij}}}/r_{ij}$, $\mathbf{r_{\mathit{ij}}}$ is the vector linking the center of
dipoles i and j, while $r_{ij}=\mid\mathbf{r}_{ij}\mid$ and $\mathds{1}$ stands for the unit dyadic tensor) and of polarizabilities matrix $\hat{\mathds{\alpha}}=diag(\bar{\bar{\boldsymbol{\alpha}}}_{1},...,\bar{\bar{\boldsymbol{\alpha}}}_{N})$ ($\bar{\bar{\boldsymbol{\alpha}}}_{i}$ being the polarizability tensor\cite{Albaladejo} associated to the $i^{th}$ object)
\begin{equation}
\bar{\bar{\boldsymbol{\chi}}}_j=\bar{\bar{\boldsymbol{\alpha}}}_{j}-i\frac{k^3}{6\pi} \bar{\bar{\boldsymbol{\alpha}}}_{j}\bar{\bar{\boldsymbol{\alpha}}}_{j}^{\dagger},
\end{equation}
\begin{equation}
\mathds{A}_{ij}=\left[\mathds{1}-k^2\hat{\mathds{\alpha}}\mathds{B}\right]_{ij}^{-1},
\end{equation}
with $\mathds{B}_{ij}=(1-\delta_{ij})\bar{\bar{\boldsymbol{G}}}_{ij}^{0}$ and
\begin{equation}
\mathds{C}_{ij}=k^2\bar{\bar{\boldsymbol{G}}}_{ik}^{0}\mathds{A}_{kj}.
\end{equation}

\begin{figure}
\includegraphics[scale=0.3]{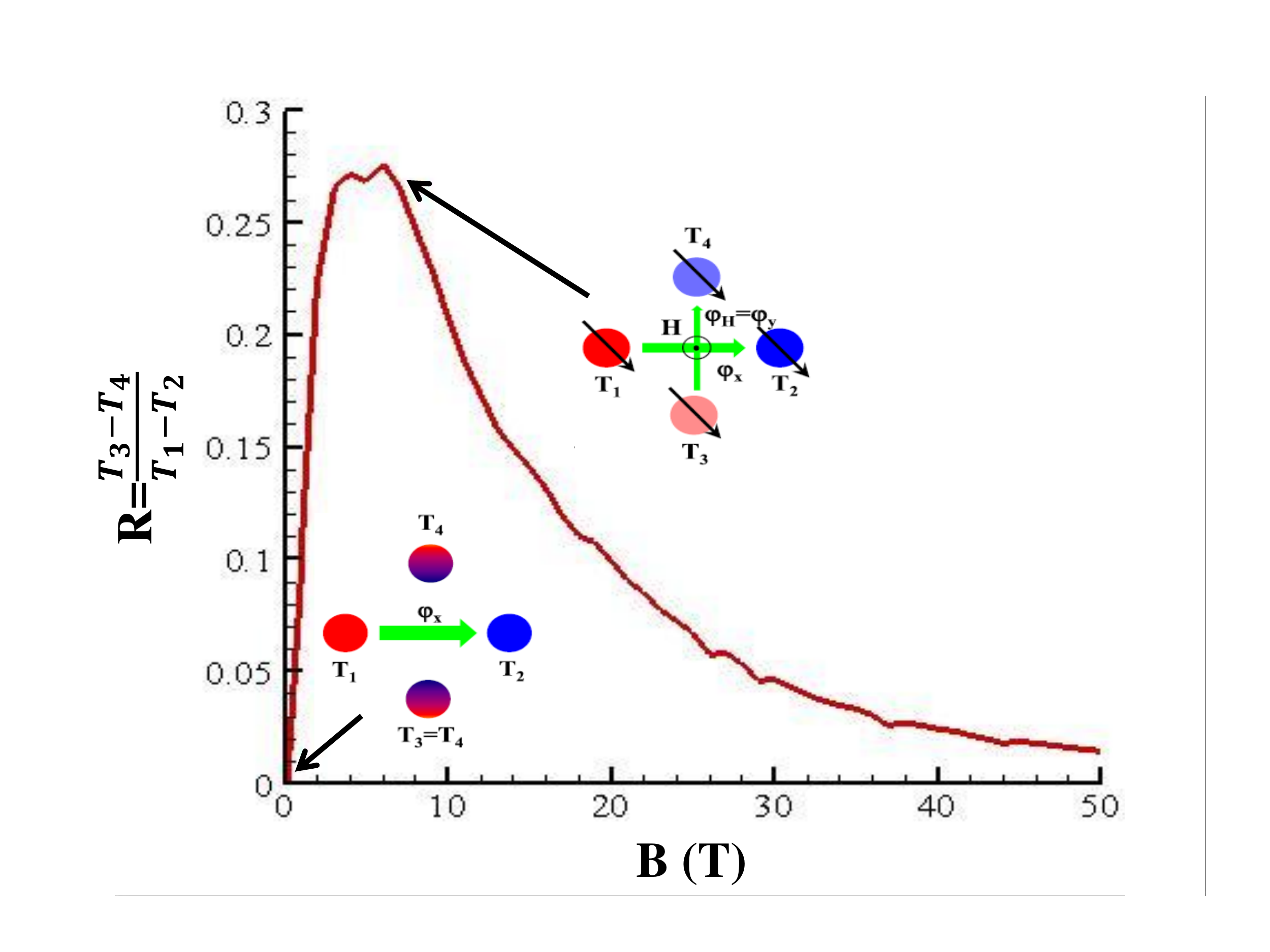}
\caption{Hall temperature difference $R$ versus the magnetic field $B$ in a four-terminal square junction of InSb particles $r=100 nm$ radius at $T_{eq}=300 K$. The separation distance $d_{12}$ (from edge to edge) is $3 r$ (i.e. near-field regime). } 
\label{Hall temperature}
\end{figure}

In Fig.~\ref{Hall temperature} we show the relative Hall temperature difference 
\begin{equation}
  R=\frac{T_3-T_4}{T_1-T_2}
\end{equation}
in a InSb-particle system~\cite{Palik} with respect to the magnitude $B$ of magnetic field  for a separation distance $d_{12}=300\,{\rm nm}$  (i.e.\ near-field regime) when the particles of radius $r=100\,{\rm nm}$ are surrounded by vacuum. When the magnetic field is zero, all particles are isotropic so that the system is symmetric and, as expected, $R=0$. On the contrary, for non-zero magnetic field the symmetry of system is broken and a Hall flux $J_H$ appears (Fig.~\ref{Hall effect}). 

\begin{figure}
\includegraphics[scale=0.3]{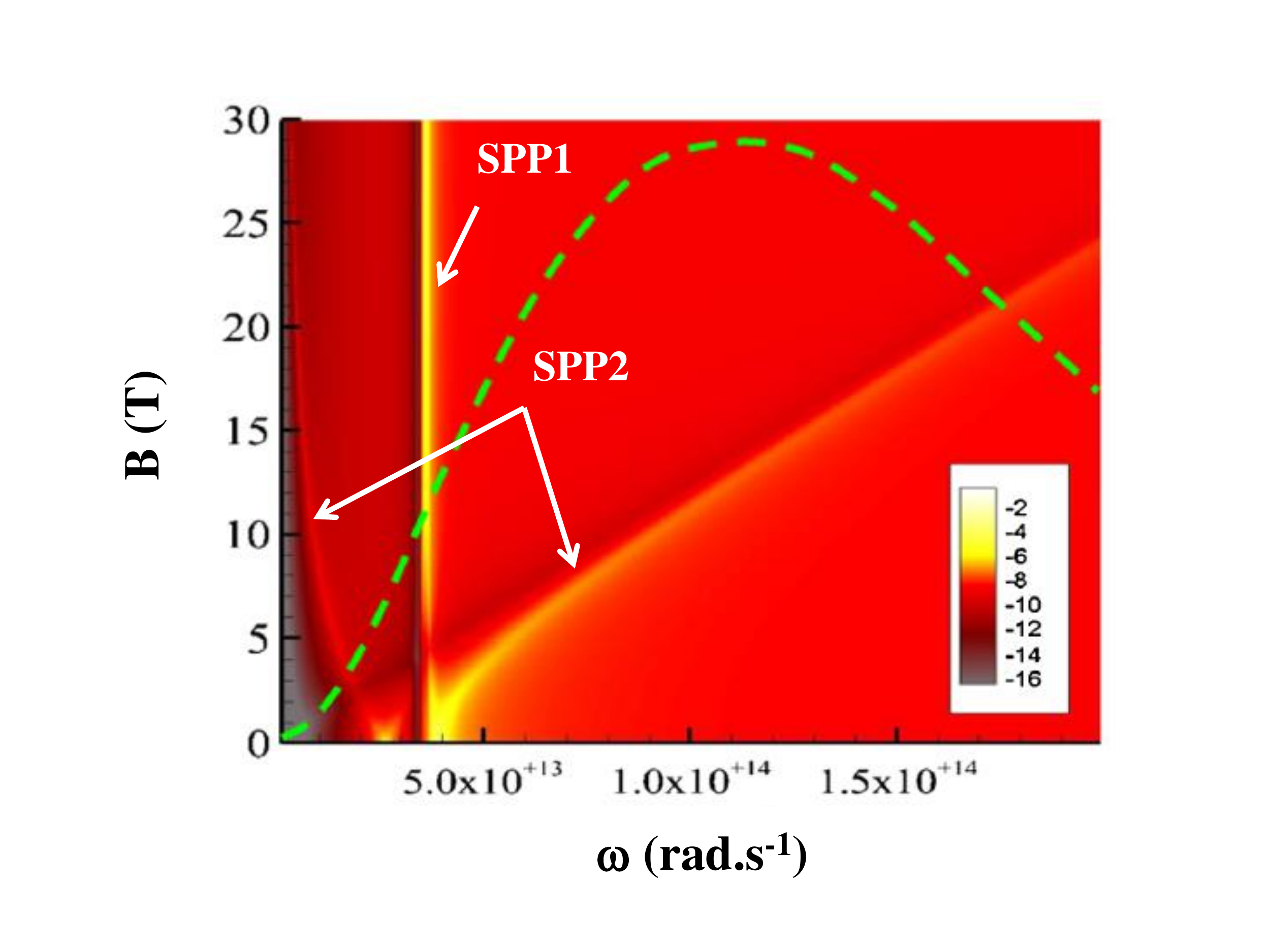}
\caption{Optical resonances (surface polaritons represented by the bright branches) of  InSb particles $r=100 nm$ radius embedded in vacuum (see~\cite{PBAHall} for more details) in the $(\omega,B)$ plane. The dashed line corresponds to the Planck function (arbitrary unit) at $T_{eq}=300 K$. } 
\label{Resonances}
\end{figure}

Under the action of the magnetic field, the spatial distribution of electric field radiated by the particles leads to a strongest dissipation of energy in the lower particle (particle 3) so that a Hall flux flows in the direction of positive $y$.  However, the magnitude of Hall flux is maximal for a magnetic field of about $B\approx6T$ (here $T$ stands vor the unit Tesla and not for the temperature). This behavior can be understood by analyzing the resonances (localized surface modes). These modes are plotted in Fig.~\ref{Resonances} with respect to the magnitude of the magnetic field.  We clearly see the presence of three different branches (bright areas). The vertical branch is independent of the magnetic field and  it corresponds to the presence of a magnetic independent surface phonon polariton (SPhP) at $\omega\sim 3.5\times10^{13}\,{\rm rad/s}$. The two others branches are two magneto-dependent surface modes. Contrary to the first resonance, these resonances  are of plasmonic nature (see~\cite{PBAHall} for more details). When the magnitude of the magnetic field becomes sufficiently large, these plasmonic resonances are shifted away from the Wien's frequency so that they do not contribute anymore to  heat exchanges. On the other hand, for weak magnetic fields, these resonances give rise to supplementary channels for heat exchanges which superimpose to the channel associated with the SPhP.  Moreover, the contribution of the high frequency plasmonic channel becomes more and more important as its frequency brings closer from the Wien's frequency. The optimal transfer occurs for a magnetic field of magnitude $B\approx6 T$.  This situation corresponds precisely to the condition where the Hall effect is maximal.

%
%

\section{Conclusion}
In this paper we have summarized the very recent advances made in optics to manipulate, amplify and even store the thermal energy with the thermal photons exchanged through various many-body systems. We have demonstrated the feasability for contactless thermal analogs of transistors, volatile memories and logic gates. These devices  allow for a management of heat flows either at macroscale (far-field exchanges) or at subwavelength scales (near-field exchanges) in complex architectures. Beside these building blocks for ``thermotronics'' we have introduced the basic theoretical framework for ``magneto-thermo-plasmonics'' which allows for controlling actively the heat flows in complex plasmonic networks using an external magnetic field. These results could find broad applications in MEMS/NEMS technologies, to generate mechanical work by using microresonators coupled to a transistor as well as in energy storage technology, for instance, to store 
and release thermal energy upon request. They could also be used to develop purely thermal wireless sensors which work by implementing logic functions with heat instead of electricity. Further analysis of the performance of such devices taking the temperature distribution inside the different parts of the system into account as shown in Ref.~\cite{MessinaEtAl2016} is now in reach.

Beyond these practical applications, the physics involved in strongly correlated many-body systems is very rich. Their study has revealed the presence of various singular behaviours. Hence, we have seen that these systems can be multistable, they can locally display negative differential resistances or, in the particular case of magneto-optic systems, they can transport a photon thermal Hall flux in presence of a magnetic field. Physics of many-body systems remains largely unexplored and there is no doubt that future research in this area will reserve surprises.

\end{document}